\documentclass[journal]{IEEEtran}
\usepackage{caption}
\usepackage{ifpdf}
\usepackage[pdftex]{graphicx}
\usepackage{cite}
\usepackage[ruled,vlined]{algorithm2e}
\usepackage{amsmath,amssymb,amsthm}
\usepackage{hyperref}
\usepackage{mathtools}
\usepackage{algorithmic}
\usepackage{makecell}
\usepackage{array}
\usepackage{booktabs,longtable}
\usepackage{array}
\usepackage{tikz}
\usepackage{lettrine}
\usepackage{tabularx}
\usepackage[caption=false,font=footnotesize]{subfig}
\usepackage{stfloats}
\usepackage{url}
\usepackage{lmodern}
\usepackage{svg}
\usepackage{esvect}
\usepackage{subcaption}
\usepackage{booktabs}
\usepackage{multirow}
\usepackage{float}
\usepackage{listings}
\usepackage{todonotes}
\usepackage{soul}   
\usepackage{xcolor}
\usepackage{amsthm}
\usepackage{varwidth}
\usepackage{placeins}
\usepackage{siunitx}
\usepackage{parskip}
\newtheorem{prob}{Problem}

\begin{document}
\title{A Hybrid Optimization Framework for \\ Spatial Packaging of Interconnected Systems}

\author{S.~Westerhof, T.~Hofman 
\thanks{S. Westerhof and T. Hofman (e-mail: t.hofman@tue.nl) are with the Eindhoven University of Technology (TU/e), Dept. of Mechanical Engineering, \href{https://www.tue.nl/en/research/research-groups/control-systems-technology}{Control Systems Technology} section, \href{https://www.tue.nl/en/research/research-groups/group-hofman}{Engineering Systems Design} group, P.O.Box 513, 5600 MB Eindhoven, The Netherlands.}}

\maketitle
\begin{abstract}
This paper presents an optimization framework for Spatial Packaging of Interconnected Systems with Physical Interactions (SPI2) that addresses the geometric challenges of three-dimensional component placement and routing. While SPI2 generally includes physical interactions, this study isolates the spatial optimization aspect to evaluate placement and routing performance independently. The framework integrates the Maximal Disjoint Ball Decomposition (MDBD) for geometric abstraction with a hybrid optimization strategy that combines stochastic initialization and gradient-based refinement with interior point optimization. It is formulated to handle the nonlinear, non-convex, and continuous characteristics of spatially coupled design problems. The proposed framework is evaluated against a use case from prior SPI2 research and tested with a newly introduced benchmark that enables verifiable assessment of optimization performance. Results indicate that the presented method achieves more than a 10\% improvement over existing SPI2 implementations and converges to spatially analytical optima across various benchmark scenarios. Benchmark experiments show solution accuracy of 0.6–2\% relative to the ground truth.
\end{abstract}

\begin{IEEEkeywords}
Spatial Packaging of Interconnected Systems with Physical Interactions (SPI2), placement and routing optimization, Maximal Disjoint Ball Decomposition (MDBD), hybrid optimization, generative design.
\end{IEEEkeywords}

\section{Introduction}\label{section:introduction}
\begin{figure}[t]
\centering
\includegraphics[width=9cm]{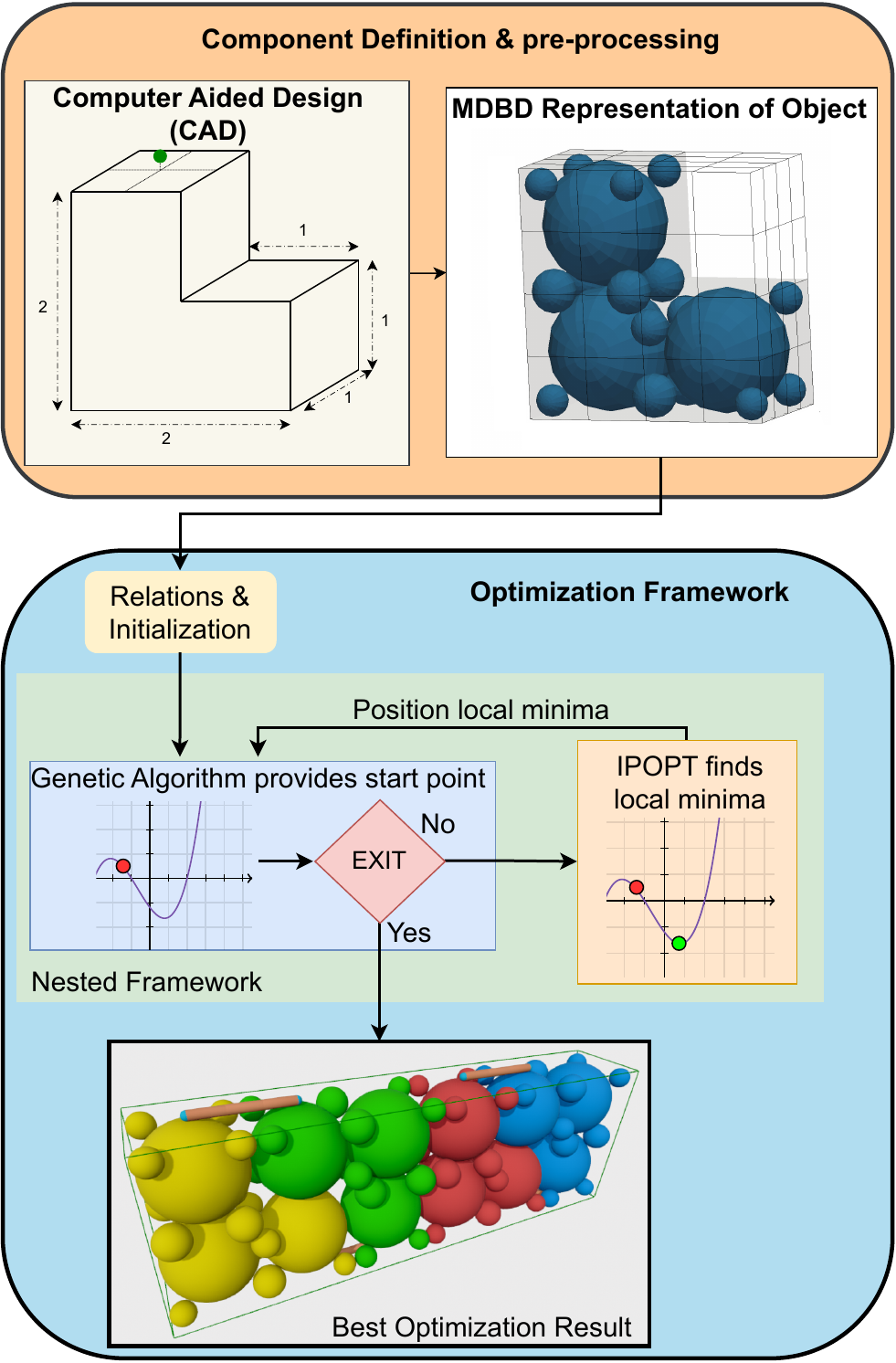}
\caption{Overview of the framework. Where a CAD design is transformed into a Maximal Disjoint Ball Decomposition (MDBD) object. After which, relations are defined. The nested framework finds a result which is output.}
\label{fig:1stpage}
\end{figure}
\lettrine{S}maller, cheaper, and more efficient — the demand for engineering systems that meet these requirements while handling complex physical interactions is growing across many applications: minimally invasive medical devices \cite{MinInv}, magnetically actuated surgical tools \cite{MagnoSur}, automotive component packaging \cite{JvKampen}\cite{Hofstetter}, commercial aircraft turbofan design \cite{Turbofan}, and military avionics \cite{Howard}\cite{Howard2}. Improving these systems reduces material use and energy consumption while increasing performance. Computational approaches such as generative design, with an emphasis on spatial placement, are therefore becoming essential.

Generative design, which uses optimization algorithms and artificial intelligence to explore design choices, offers a transformative approach to this challenge \cite{Borsboom}. Rather than serving as a magic bullet, it extends the design process — much like finite element analysis for structural design or computational fluid dynamics for aerodynamics. Generative design automates the exploration and comparison of complex alternatives, overcoming the limits of human decision-making. Instead of relying solely on designer intuition, it systematically evaluates a large number of configurations to balance competing needs \cite{IntroSPI2}. Yet, in highly integrated systems where spatial placements interact, an additional optimization layer is required.

Spatial Packaging of Interconnected Systems with Physical Interactions (SPI2) frameworks \cite{Behzadi} evaluate the system as a whole. By modeling mechanical, thermal, and electrical dependencies together, SPI2 could enable faster discovery of compact layouts compared to manual iterations. In essence, SPI2 treats the spatial arrangement and interconnections of components as a coupled optimization problem, where the geometry, physical behavior, and routing constraints are solved simultaneously rather than in isolation.

Three challenges in current SPI2 frameworks motivate the present work. First, existing SPI2 implementations rely on a limited set of optimization strategies, making it unclear whether alternative frameworks could improve convergence speed or robustness in complex, nonconvex design spaces.  Second, there is no established way to determine what constitutes a good or near-optimal result. Since both the spatial model and the optimization algorithm jointly shape the loss landscape, their combined accuracy must be evaluated against a known ground truth—an aspect currently missing in the literature. Third, the inherent NP-hardness of spatial placement and routing causes combinatorial growth in computational complexity, severely limiting scalability \cite{NPhardPipe, NPhardPlacement}. Addressing this requires effective decomposition strategies and constraint-reduction methods that maintain feasibility while reducing computational load.

Against this backdrop, this work introduces a novel placement and routing optimization framework, together with a benchmark use case, to enable quantitative evaluation and verification of optimization effectiveness. Additionally, we investigate possible solution directions to decompose the placement and routing problem.

\subsubsection*{Related Literature}
This investigation relates to two main research streams: placement and routing problems, and optimization methodology.

The first research stream, placement and routing problems, can be divided into two categories: geometric representation and SPI2-related research.

Geometric representation: CAD-based design tools typically describe 3D geometries using point clouds, surface meshes \cite{STL}, or NURBS curves \cite{Schneider}. While accurate for modeling and visualization, these representations are computationally intensive and unsuitable for gradient-based optimization because of their discrete nature. To address this, simplified geometric abstractions are used. The Geometric Projection Method (GPM) \cite{Bello, Norato} approximates objects using enclosing spheres or ellipsoids, enabling fast computation but often leaving unused volumes. Alternatively, the Maximal Disjoint Ball Decomposition (MDBD) method \cite{Behzadi, Chen} fills the object with the largest non-overlapping spheres, achieving higher geometric fidelity while supporting gradient-based optimization. The number of spheres can be tuned to balance accuracy and computational cost. In this study, the MDBD method is adopted.

SPI2-related research: Two main research groups have recently advanced SPI2 frameworks. The most significant difference between them lies in how components are represented. The group led by Allison at the University of Illinois Urbana–Champaign employs the \textit{Geometric Projection Method} (GPM) \cite{IntroSPI2, Bello, Underhood, Simultaneous, 3DGeometric, ANovel, MachineLearning, Systematic, Machine}. In contrast, the group led by Ilieş at the University of Connecticut applies the \textit{Maximal Disjoint Ball Decomposition} (MDBD) method \cite{Behzadi, Chen, Chen2, Vu}. These groups have also collaborated to identify current gaps and challenges in SPI2 optimization \cite{Holistic}. Other relevant contributions include 2D powertrain layout optimization \cite{JvKampen} and an investigation of air-compressor system optimization \cite{aircompressor}.

The second research stream, optimization methodology, focuses on approaches suitable for constrained, nonlinear, non-convex, and continuous optimization problems. This stream directly complements the SPI2 literature, as the choice of optimization method influences the placement–routing formulation.

Current SPI2 frameworks \cite{Bello, Underhood, Simultaneous, Machine, ANovel, Behzadi, ModelBased} share a crucial limitation: they typically assess performance by comparing only the initial and final configurations \cite{Bello}, often reporting the percentage improvement in the objective function. Yet, spatial placement, routing, and physics form a constrained, nonlinear, and non-convex optimization problem \cite{Underhood, 3DGeometric, Simultaneous, ANovel, Behzadi}. Because of this complexity, such comparisons provide no indication of whether the obtained solution is near the true or analytical optimum. As a result, while reported improvements demonstrate clear progress, they are difficult to interpret in terms of absolute solution quality. Establishing verifiable benchmarks would therefore strengthen the reliability and generalizability of SPI2 frameworks and enhance their value for both scientific study and practical design applications.

According to \cite{Holistic}, SPI2 optimization typically proceeds in three stages: initialization, topology selection, and integrated placement–routing–physics optimization. The present work focuses exclusively on the placement and routing stage. Both the placement and routing problems are individually NP-hard \cite{NPhardPipe, NPhardPlacement}. To improve the likelihood of finding feasible solutions, the placement–routing–physics optimization is formulated as a constrained, nonlinear, and non-convex continuous optimization problem \cite{Underhood, 3DGeometric, Simultaneous, ANovel, Behzadi}. Gradient-based solvers can find solutions that might otherwise remain undiscovered. Group Allison typically uses the \textit{Method of Moving Asymptotes} (MMA), a nonlinear programming approach developed by Svanberg \cite{Norato, Svanberg}. Other approaches include MATLAB’s \textit{fmincon} \cite{fmincon} with an active-set algorithm, which was used in \cite{3DGeometric}. Another investigation \cite{Behzadi} reformulates the constrained problem into an unconstrained one, enabling the use of the stochastic gradient descent (SGD) with the automatic differentiation method ADAM \cite{ADAM}.

These existing implementations rely on a narrow set of optimization strategies; there is an opportunity to explore alternative algorithms capable of solving constrained, nonlinear, and non-convex continuous optimization problems within SPI2 frameworks.

Although current SPI2 studies demonstrate promising improvements, it remains unclear how close these solutions are to the true spatial optimum, as no prior work has evaluated SPI2 optimization frameworks against configurations with analytically known solutions. This lack of verification limits trust in the obtained layouts, especially in practical design settings where the true optimum is unknown. Therefore, a central motivation of this work is to evaluate how the proposed optimization framework compares with existing SPI2 solvers and to verify whether it reliably converges to known analytical optima when they exist. Establishing this reliability is essential, as the ultimate goal is to deploy the framework in real design applications where ground-truth solutions are unavailable and the optimization must be trusted to deliver accurate and consistent results.


\subsubsection*{Statement of Contributions}
This paper addresses the three challenges identified in the Background by proposing, validating, and analyzing a framework for spatial placement and routing within SPI2. Concretely, we contribute: a nested hybrid optimization framework that couples stochastic exploration (random and genetic initializations) with deterministic gradient-based refinement using IPOPT via CasADi, as shown in Fig.~\ref{fig:1stpage}. We introduce three verifiable benchmarks with analytically known optima, enabling quantitative accuracy checks of the combined model and optimizer. A comparison study is performed against the MDBD-based SGD baseline of \cite{Behzadi}, allowing direct evaluation of relative performance in the combined placement–routing objective. A scalability study is conducted to measure interior-point iteration time as a function of the number of objects, spheres, and control points, providing insights into computational growth. A framework for spatial decomposition via Analytical Target Cascading (ATC) and Sphere of Influence (SOI) is implemented, enabling assessment of different decomposition strategies for placement–routing problems. Constraint-reduction tactics are examined, comparing Soft-Sum and Absolute formulations to evaluate their impact on computational efficiency and feasibility. Finally, a systematic initialization study explores the influence of initialization methods (Equally Spaced, Genetic Algorithm, and Random) and investigates warm-starting from lower- to higher-resolution sphere counts.

\subsubsection*{Organization}
The remainder of this paper is structured as follows: Section~\ref{section:Methods} introduces the model, optimization objectives, and frameworks. Section~\ref{section:use-case} introduces the use-case for a direct comparison against \cite{Behzadi} and novel use-cases on which the framework output is verifiable. The results are presented in Section~\ref{section:results}. Next, Section~\ref{section:conclusion} provides a conclusion, and Section~\ref{sec:Future} presents a direction for future research.

\section{Methods}\label{section:Methods}
\subsection{Maximal Disjoint Ball Decomposition}
\begin{figure}[t]
    \centering
    \includegraphics[width=9cm]{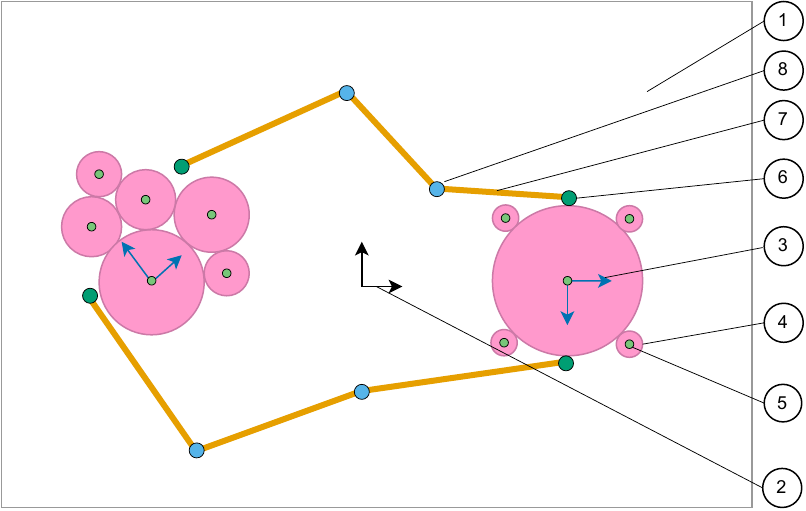}
    \caption{Depiction of the model with various components: \(n_\mathrm{d}\)-dimensional workspace \(\mathbb{W}\) (1, White), Cartesian frame \(\mathbb{F}_\mathbb{W}\) (2, Black), Object Frame \(\mathbb{F}_{A_i}\) (3, Dark Blue), Spheres \(b_{i,\mu}\) (4, Light Purple), center \(\mathbf{p}_{\mathrm{b}_{i,\mu}}\) (5, Light Green), ports \(\varphi_{i,\ell}\) (6, Dark Green), Routing segments (7, Orange), control points \(\mathbf{c}_{L,k}\) (8, Light Blue)}
    \label{fig:model}
\end{figure}

Previous research has examined the mathematical formulation and geometric properties of Maximal Disjoint Ball Decomposition (MDBD) representations~\cite{Vu, Chen, Chen2}. Following the definition in~\cite{Chen}, a disjoint spherical decomposition is a collection of non-overlapping, closed \( n_\mathrm{d} \)-dimensional balls, where \( n_\mathrm{d} \) denotes the number of spatial dimensions. An MDBD object \( A_i \) is therefore defined as
\begin{equation}
   A_i = \{\, b_{i,1}, b_{i,2}, \dots, b_{i,\,n_{\mathrm{b},i}} \,\},
\end{equation}
representing the set of spheres that together approximate the geometry of component \( i \).
Each sphere \( b_{i,\mu} \) is defined by its center position \( \mathbf{p}_{\mathrm{b}_{i,\mu}} \) and radius \( r_{\mathrm{b}_{i,\mu}} \), and all spheres within the same object are mutually disjoint, i.e., \( b_{i,\mu} \cap b_{i,\nu} = \emptyset \) for all \( \mu \neq \nu \).
The parameter \( n_{\mathrm{b},i} \) specifies the total number of spheres used to represent object \( i \).
Consequently, each decomposition can be uniquely described by the set of sphere centers and radii.

This formalization provides a mathematical foundation for MDBD as a compact, geometric representation suitable for gradient-based optimization. The method for creating such objects can be found in Appendix \ref{Appendix:A}. 

\subsection{Model Setup}\label{ModelSetup}
The setup of the model is very similar to Behzadi et al. And for this section, almost the complete setup from \cite{Behzadi} is utilized with minor differences to suit the program used in this paper. Where the method differs from \cite{Behzadi} will be explicitly stated.

Consider an \(n_\mathrm{d}\)-dimensional workspace \(\mathbb{W}\) (1, White) with Cartesian frame \(\mathbb{F}_\mathbb{W}\) (2, Black) as seen in Fig.~\ref{fig:model}. We have \(n_\mathrm{obj}\) rigid objects. Object \(A_i\) has its own frame \(\mathbb{F}_{A_i}\) (3, Dark Blue) and consists of a set of \(n_{\mathrm{b},i}\) spheres. Sphere \(b_{i,\mu}\) (4, Light Purple) has center \(\mathbf{p}_{\mathrm{b}_{i,\mu}}\) (5, Light Green) and radius \(r_{\mathrm{b}_{i,\mu}}\). Now a minor difference from Behzadi et al, it that each object may also have ports \(\varphi_{i,\ell}\) (6, Dark Green) with \(\ell=1,\dots,n_{\varphi,i}\), instead of being connected to the center of an object. \(n_{\varphi,i}\) is the total number of ports in object $i$. Routing segments (7, Orange) span linearly between ports and control points \(\mathbf{c}_{L,k}\) (8, Light Blue). Each route $L$ consists of $K_L$ straight segments that connect $K_L + 1$ nodes \([\mathbf{q}_{L,0},\mathbf{q}_{L,1},\dots,\mathbf{q}_{L,K_L}]\), where the endpoints \(\mathbf{q}_{L,0}\) and \(\mathbf{q}_{L,K_L}\) are ports and the intermediate nodes \(\mathbf{q}_{L,k}(k=1,\dots,K_L - 1)\) are control points \(\mathbf{c}_{L,k}\). The number of segments is \(K_L = n_{\mathrm{cp},L}+1\), and segments \(m\) connect nodes \(\mathbf{q}_{L,m}\) and \(\mathbf{q}_{L,m+1}\) for \(m=0,\dots,K_L-1\). Where \(n_{\mathrm{cp},L}\) is the total number of control points across a single route \(L\). The routing tube has constant radius \(r_r\).

From now on, \(n_\mathrm{d}=3\). 
Each object position is parameterized by
\begin{equation}
    \mathbf{x}_{\mathrm{A}_i}=\big[x_{\mathrm{\theta},i},\,x_{\mathrm{\alpha},i},\,x_{\mathrm{\beta},i},\,x_{\mathrm{x},i},\,x_{\mathrm{y},i},\,x_{\mathrm{z},i}\big]^\top\in\mathbb{R}^6,
\end{equation}
where \((x_{\mathrm{\theta},i},x_{\mathrm{\alpha},i},x_{\mathrm{\beta},i})\) are angles in radians around the $\mathrm{x}$, $\mathrm{y}$, and $\mathrm{z}$ axis, and \((x_{\mathrm{x},i},x_{\mathrm{y},i},x_{\mathrm{z},i})\) are translations corresponding the the cartesian reference frame, with arbitrary units. 
Control points do not have an angular component and can be represented as
\begin{equation}
    \mathbf{c}_{L,k}=[c_{\mathrm{x},L,k},\,c_{\mathrm{y},L,k},\,c_{\mathrm{z},L,k}]^\top\in\mathbb{R}^3.
\end{equation}

A rigid transform from the object frame \(\mathbb{F}_{A_i}\) to the workspace frame \(\mathbb{F}_\mathbb{W}\) is
\begin{equation}
    \mathbf{p}^{\mathbb{W}}=\mathbf{R}_i\,\mathbf{p}^{A_i}+\mathbf{t}_i,
\quad
\mathbf{R}_i\in\mathbb{R}^{3\times 3},\ \ \mathbf{t}_i\in\mathbb{R}^{3},
\end{equation} where \(\mathbf{p}^{A_i}\) is a center point of either a sphere \(\mathbf{p}_{\mathrm{b}_{i,\mu}}\), or a node location \(\mathbf{q}_{L,K_L}\). And \(\mathbf{t}_i\) is a translation vector which is either from the object design variables \((x_{\mathrm{x},i},x_{\mathrm{y},i},x_{\mathrm{z},i})\) or the control point design variable \(\mathbf{c}_{L,k}\).

We use Roll–Pitch–Yaw angles (elementary rotations about a single frame's axes) \cite{Multibod}; the rotation matrix is \(\mathbf{R}_i=\mathbf{R}_z(x_{\mathrm{\theta},i})\,\mathbf{R}_y(x_{\mathrm{\alpha},i})\,\mathbf{R}_x(x_{\mathrm{\beta},i})\). With the first object fixed to the workspace, the total number of design variables is
\begin{equation}
    n_\mathrm{var}=6\,(n_\mathrm{obj}-1)+3\sum_L n_{\mathrm{cp},L},
\end{equation}
and the design variables become
\begin{equation}
    \mathbf{x}=\big[\ \mathbf{x}_2^\top,\dots,\mathbf{x}_{n_\mathrm{obj}}^\top,\ \mathbf{c}_{1,1}^\top,\dots,\mathbf{c}_{L,k}^\top\ \big]^\top.
\end{equation}

\subsection{Constraint Equations}
Now, the objects are not allowed to interfere with each-other, the routing cannot interfere with the objects, and the routing cannot interfere with itself. This produces three types of constraints; object-object interference, routing-object interference, and routing-routing interference. This subsection further elaborates on how these are constructed. The methodology of the constraint equations is very similar to the constraints constructed by \cite{Chen2}. Similar definitions of these constraint equations are also used in \cite{Behzadi}.

\subsubsection{Object-Object Interference}
One advantage of using spheres for geometric representation is that interference checks are straightforward to compute, even though the overall problem remains non-convex. 
Each sphere \(b_{i,\mu}\) has center \(\mathbf{p}^{\mathbb{W}}_{\mathrm{b}_{i,\mu}}\in\mathbb{R}^3\) and radius \(r_{\mathrm{b}_{i,\mu}}\).

For every unique object pair \((i,j)\) with \(i<j\), where \(j\) is an object index, and for every sphere pair
\(\mu=1,\dots,n_{\mathrm{b},i}\) and \(\nu=1,\dots,n_{\mathrm{b},j}\), the distance between the spheres can be found as
\begin{equation}
  d^{\,\mathrm{obj\mbox{-}obj}}_{i,\mu,j,\nu}
  \;=\;
  \left\|\,\mathbf{p}^{\mathbb{W}}_{\mathrm{b}_{i,\mu}}
           -\mathbf{p}^{\mathbb{W}}_{\mathrm{b}_{j,\nu}}\,\right\|_2
  \;-\;
  \bigl(r_{\mathrm{b}_{i,\mu}}+r_{\mathrm{b}_{j,\nu}}\bigr).
  \label{eq:clearance_objobj}
\end{equation}

The spheres do not overlap when
\begin{equation}
  d^{\,\mathrm{obj\mbox{-}obj}}_{i,\mu,j,\nu}\;\ge\;0
  \qquad
  \forall\, i<j,\;\mu=1,\dots,n_{\mathrm{b},i},\;\nu=1,\dots,n_{\mathrm{b},j}.
  \label{eq:nonoverlap_objobj}
\end{equation}

To include all pair-wise sphere-to-sphere distances, we collect them as
\begin{equation}
  \mathbf{g}^{\mathrm{obj\mbox{-}obj}}
  \;=\;
  -\bigl[\,d^{\,\mathrm{obj\mbox{-}obj}}_{i,\mu,j,\nu}\,\bigr]_{(i,j,\mu,\nu)}
  \;\in\;\mathbb{R}^{N_{\mathrm{pairs}}},
\end{equation}
where the total number of sphere-pair constraints is
\begin{equation}
  N_{\mathrm{pairs}}
  \;=\;
  \sum_{1\le i<j\le n_\mathrm{obj}} n_{\mathrm{b},i}\,n_{\mathrm{b},j}.
\end{equation}

Finally, these constraints are enforced in negative-null form as
\begin{equation}
    \mathbf{g}^{\mathrm{obj\mbox{-}obj}}(\mathbf{x}) \;\leq\; 0.
\end{equation}

\subsubsection{Routing-Object Interference}
The routing spans between ports and control points, as illustrated in Fig.~\ref{fig:model}. The shortest distance from each routing segment to the surrounding spheres can be computed. The shortest distance between the center of the segment and a sphere center is denoted \(d_{L,m,i,\mu}\). Additionally, the constant radius \(r_\mathrm{r}\) of the routing tube and the sphere radius \(r_{\mathrm{b}_{i,\mu}}\) the distance relation becomes
\begin{equation}
    d^{\,\mathrm{route\mbox{-}obj}}_{L,m,i,\mu}
    = \bigl\|\mathbf{p}^{\mathbb{W}}_{\mathrm{b}_{i,\mu}}-\mathbf{p}_{\mathrm{proj},L,m}\bigr\|_2
      - \bigl(r_{\mathrm{b}_{i,\mu}} + r_r\bigr),
\end{equation}
where \(\mathbf{p}_{\mathrm{proj},L,m}\) is the closest point on the segment to the sphere center. To avoid interference \(d^{\,\mathrm{route\mbox{-}obj}}_{L,m,i,\mu} \;\ge\; 0\) must be true. 
This holds for all routes \(L\). Stacking all routing–object clearances yields
\begin{equation}
    \mathbf{g}^{\mathrm{route\text{-}obj}}
= -\bigl[\,d^{\,\mathrm{route\mbox{-}obj}}_{L,m,i,\mu}\,\bigr]_{(L,m,i,\mu)}
\in \mathbb{R}^{N_{\mathrm{r\text{-}o}}},
\end{equation}

with the number of constraint equations
\begin{equation}
N_{\mathrm{r\text{-}o}} =
\sum_{L}\sum^{K_L-1}_{m=0}\sum_{i=1}^{n_\mathrm{obj}} n_{\mathrm{b},i}.
\end{equation} And to be used in negative-null form as
\begin{equation}
    \mathbf{g}^{\mathrm{route\text{-}obj}}(\mathbf{x})\leq 0.
\end{equation} A more elaborate explanation of the distance calculation can be found in Appendix \ref{Appendix:B}.

\subsubsection{Routing–Routing Interference}
Routing tubes must not intersect each other. Each route \(L\) consists of \(K_L\) segments, where \(m\) connects the nodes \(\mathbf{q}_{L,m}\) and \(\mathbf{q}_{L,m+1}\) for \(m=0,\dots,K_L-1\). To evaluate interference between routes, consider a second route \((L')\) with its own segments \(\eta=0,\dots,K_L-1\). The shortest distance between any two segments \((L,m)\) and \((L',\eta)\) defines the clearance, from which the combined tube radii \(2r_r\) are subtracted.

To avoid double counting and to allow adjacent segments of the same route to meet at their shared endpoint, we only check for pairs in the set
\begin{equation}
\begin{aligned}
\mathcal{C}
=\Bigl\{(L,m,L',\eta)\ \big|\
\big(L<L',\ 0\le m<K_L,\ 0\le \eta<K_{L'}\big)\\
\text{or}\ 
\big(L=L',\ 0\le m<\eta<K_L,\ |m-\eta|\ge 2\big)
\Bigr\}.
\end{aligned}
\label{eq:SetS}
\end{equation}
Put simply, the set $\mathcal{C}$ tells us which pairs of the routing segments we need to check for collisions. For different routing, check every pair once. For segments in the same routing, only check non-adjacent pairs (so segments sharing a node can meet). This ensures we cover all possible interfaces without redundant or invalid comparisons.

For all \((L,m,L',\eta)\in\mathcal{C}\), let \(d^{\,\mathrm{route\mbox{-}route}}_{L,m,L',\eta}\) be the clearance
(see Appendix~\ref{Appendix:B} for the exact formula). The non-interference condition is
\begin{equation}
    d^{\,\mathrm{route\mbox{-}route}}_{L,m,L',\eta}\ \ge\ 0
\quad\ \forall\ (L,m,L',\eta)\in\mathcal{C}.
\end{equation}
Stacking all constraints gives
\begin{equation}
    \mathbf{g}^{\mathrm{route\mbox{-}route}}
= -\bigl[d^{\,\mathrm{route\mbox{-}route}}_{L,m,L',\eta}\bigr]_{(L,m,L',\eta)\in\mathcal{C}}
\in \mathbb{R}^{N_{\mathrm{r\mbox{-}r}}},
\end{equation}
where \(N_{\mathrm{r\mbox{-}r}} = \lvert\mathcal{C}\rvert\), and we enforce
\begin{equation}
\mathbf{g}^{\mathrm{route\mbox{-}route}}(\mathbf{x})\ \le\ 0
\end{equation}

The total number of constraints \(N_{\mathrm{tot}}=N_{\mathrm{pairs}}+N_{\mathrm{r\text{-}o}}+N_{\mathrm{r\mbox{-}r}}\) grows combinatorial with the number of objects, spheres, and routing segments, which has practical implications for solver scalability.

%
\subsection{Intermezzo: Smoothing Function}
Interior-point methods with automatic differentiation benefit from smooth, everywhere-differentiable functions. We therefore use the Boltzmann (soft) operator \cite{Wiki}, following \cite{Chen2}:
\begin{equation}
S_{\alpha}(x_1,\dots,x_n)=
\frac{\sum_{i=1}^{n} x_i\, e^{\alpha x_i}}{\sum_{i=1}^{n} e^{\alpha x_i}}.
\end{equation}
The function approximates a maximum or minimum function depending on whether the value of \(\alpha\) is positive or negative. If \(\alpha\) approaches \(\infty\), then \(\mathcal{S}_{\alpha}(x_1,\dots,x_n)\) approaches the \(\mathrm{max}\) operator, and vice versa for \(-\infty\). The variable \(x\) in this case represents an input variable used to determine the maximum or minimum value. There are \(n\) input variables.

\subsection{Soft-Sum-Constraints}
In the previous sections, the constraint equations were defined per sphere or per routing segment. 
However, this leads to combinatorial growth in constraint equations when the number of objects, spheres, and/or routing segments increase.  
An alternative is to take a smooth maximum over these constraints:
\begin{equation}
    g_\mathrm{soft}^{\mathrm{obj\mbox{-}obj}} 
    = \mathop{S_{\alpha}}\limits_{\forall\, \mathbf{g}^{\mathrm{obj\mbox{-}obj}}}
         (\mathbf{g}^{\mathrm{obj\mbox{-}obj}}),
\end{equation}
\begin{equation}
    g_\mathrm{soft}^{\mathrm{route\mbox{-}obj}} 
    = \mathop{S_{\alpha}}\limits_{\forall\, \mathbf{g}^{\mathrm{route\mbox{-}obj}}}
         (\mathbf{g}^{\mathrm{route\mbox{-}obj}}),
\end{equation}
\begin{equation}
    g_\mathrm{soft}^{\mathrm{route\mbox{-}route}} 
    = \mathop{S_{\alpha}}\limits_{\forall\, \mathbf{g}^{\mathrm{route\mbox{-}route}}}
         (\mathbf{g}^{\mathrm{route\mbox{-}route}}).
\end{equation}
The advantage of this is that the number of constraints is limited to these three and therefore the optimization quicker. However, with the soft-sum constraints the interference is not strictly enforced.

\subsection{Objective Functions}
This paper tests multiple scenarios using four distinct objective functions to ensure all cases are covered.

First, the objective \(f_\mathrm{vr}\) represents the volume of an Axis-Aligned Bounding Box (AABB). An AABB is the smallest rectangular box aligned with the coordinate axes that completely encloses all objects and routing points in the workspace, providing a measure of the total occupied volume, and is defined as
\begin{align}
        f_\mathrm{vr}(\mathbf{x}) &=
\prod_{\kappa=1}^{n_d} \Biggl(
   S_\alpha\!\Bigl(
      \mathop{S_\alpha}\limits_{\forall\, b_{i,\mu}}
         (\mathbf{p}^{(\kappa)}_{\mathrm{b}_{i,\mu}}+r_{\mathrm{b}_{i,\mu}}),
      \mathop{S_\alpha}\limits_{\forall\, \mathbf{q}_{L,k}}
         (\mathbf{q}^{(\kappa)}_{L,k})
   \Bigr) \nonumber \\
&\quad - S_{-\alpha}\!\Bigl(
      \mathop{S_{-\alpha}}\limits_{\forall\, b_{i,\mu}}
         (\mathbf{p}^{(\kappa)}_{\mathrm{b}_{i,\mu}}-r_{\mathrm{b}_{i,\mu}}),
      \mathop{S_{-\alpha}}\limits_{\forall\, \mathbf{q}_{L,k}}
         (\mathbf{q}^{(\kappa)}_{L,k})
   \Bigr)
\Biggr).
\end{align}
This includes all objects and routing points, similar to \cite{Chen2} and \cite{Behzadi}. This objective function is used for an \textit{"apples-to-apples"} comparison of the use-case presented in \cite{Behzadi}. 

Second, \(f_\mathrm{v}\) is the volume of the AABB only including the objects
\begin{align}
f_\mathrm{v}(\mathbf{x}) &=
\prod_{\kappa=1}^{n_d} \Biggl(
      \mathop{S_\alpha}\limits_{\forall\, b_{i,\mu}}
         (\mathbf{p}^{(\kappa)}_{\mathrm{b}_{i,\mu}}+r_{\mathrm{b}_{i,\mu}})
      - \mathop{S_{-\alpha}}\limits_{\forall\, b_{i,\mu}}
         (\mathbf{p}^{(\kappa)}_{\mathrm{b}_{i,\mu}}-r_{\mathrm{b}_{i,\mu}})
\Biggr),
\end{align}
which is used for the new use-case which will be further elaborated on in Section~\ref{section:use-case}. 

Third, for the routing length, \(f_\mathrm{rq}\) is the total routing length squared
\begin{align}
f_\mathrm{rq}(\mathbf{x}) &=
\sum_{L}\sum_{m=0}^{K_L-1} 
   \bigl\| \mathbf{q}_{L,m+1} - \mathbf{q}_{L,m} \bigr\|_2^2,
\label{eq:frl}
\end{align}
and, fourth, \(f_\mathrm{re}\) scales exponentially with an increased routing length
\begin{equation}
    f_\mathrm{re}(\mathbf{x}) =
\sum_{L}\sum_{m=0}^{K_L-1} e^{\bigl\| \mathbf{q}_{L,m+1} - \mathbf{q}_{L,m} \bigr\|_2}-1.
\end{equation}
These formulations differ from those used in \cite{Behzadi}, where a linear objective function was adopted. In the present work, a linear objective was found unsuitable when the shortest route corresponds to a straight line between two ports. In such cases, all control points align along an affine line, causing the objective value to remain constant and the gradient to vanish. This results in a flat search space with an indeterminate optimum, which prevents CasADi \cite{CasADi} from converging to a valid solution.

\subsection{Optimization Problems}
In this section multiple optimization problems are presented, which enables various test configurations to be used by the optimization framework in various conditions and various use-cases. There are four distinct objective functions which are being used in this investigation;

\begin{align}
    f_1(\mathbf{x}) &= f_\mathrm{vr}(\mathbf{x})w_\mathrm{vr} + f_\mathrm{rq}(\mathbf{x})w_\mathrm{rq},\\
    f_2(\mathbf{x}) &= f_\mathrm{vr}(\mathbf{x})w_\mathrm{vr} + f_\mathrm{re}(\mathbf{x})w_\mathrm{re},\\
    f_3(\mathbf{x}) &= f_\mathrm{v}(\mathbf{x})w_\mathrm{v} + f_\mathrm{rq}(\mathbf{x})w_\mathrm{rq},\\
    f_4(\mathbf{x}) &= f_\mathrm{v}(\mathbf{x})w_\mathrm{v} + f_\mathrm{re}(\mathbf{x})w_\mathrm{re},
\end{align}
where \(w_\mathrm{ab}\) are the corresponding unit-less weight factors to their respective functions.

\begin{prob}\label{prob:1}
Optimization with absolute constraints
\begin{equation*}
\begin{aligned}
& \min && f_{n_\mathrm{prob}}(\mathbf{x})\\
& \text{s.t.} && \mathbf{g}^{\mathrm{obj\mbox{-}obj}}(\mathbf{x}) \leq 0,\\
& && \mathbf{g}^{\mathrm{route\text{-}obj}}(\mathbf{x})  \leq 0, \\
& && \mathbf{g}^{\mathrm{route\mbox{-}route}}(\mathbf{x})  \leq 0.
\end{aligned}
\end{equation*}
\end{prob}
Where \(n_\mathrm{prob}\) is the integer of the problem spanning between one and four. 

\begin{prob}\label{prob:2}
Optimization with Soft-Sum constraints
\begin{equation*}
\begin{aligned}
& \min && f_{n_\mathrm{prob}}(\mathbf{x})\\
& \text{s.t.} && g_\mathrm{soft}^{\mathrm{obj\mbox{-}obj}} \leq 0,\\
& && g_\mathrm{soft}^{\mathrm{route\mbox{-}obj}}  \leq 0, \\
& && g_\mathrm{soft}^{\mathrm{route\mbox{-}route}} \leq 0.
\end{aligned}
\end{equation*}
\end{prob}
Problem \ref{prob:1} and Problem \ref{prob:2} only differ in the constraint equations where the first problem enforces constraints absolutely and the second constraint enforces the constraints with the soft-sum approach. The purpose of this is to be able to test which formulation will find a quick and accurate representation. Both problems are constraint, non-linear, non-convex, and continuous optimization problems. The next sub-section will present three optimization frameworks to solve the introduced problems.

\subsection{Optimization Frameworks}
In this section, three optimization frameworks are presented to solve the placement and routing problem. As stated previously, the problems are constrained, non-linear, non-convex, and continuous; therefore, nonlinear programming (NLP) methods are used.

In this context, the interior-point optimization solver IPOPT from CasADi~\cite{CasADi, IPOPT} was adopted as the primary solver due to its robust constraint handling and Python integration. Unlike the Method of Moving Asymptotes (MMA) or ADAM, IPOPT provides Lagrange multiplier updates and second-order convergence properties, which are useful handeling a large number of constraints.

However, due to the highly non-linear nature of the placement and routing problem, the likelihood of convergence to a poor local optimum is high. To improve the chances of identifying better local optima, a \emph{nested hybrid} strategy is adopted. This combines stochastic exploration with deterministic gradient-based refinement via IPOPT. The problem is initialized multiple times, and the solver refines each initialization until convergence. By repeating this process until an exit condition is met, the framework increases the likelihood of finding the best local solution.

Given the exponential growth of computational complexity with system size, decomposition methods are also considered to enhance scalability. In particular the decomposition strategy Analytical Target Cascading (ATC)~\cite{ATC, ATC2, ATC3} are explored. ATC enables large, coupled problems to be partitioned into smaller subproblems that can be solved in parallel while maintaining system-level consistency.

Three frameworks are presented in total: a purely nested approach, a nested approach with ATC, and a nested approach combining the Geometric Projection Method (GPM) with MDBD, referred to as the \emph{Sphere of Influence} (SOI) approach. It must be noted that the full placement and routing problem is solved only with the first optimization framework, while the ATC and SOI approaches focus exclusively on the placement problem.

Before discussing these frameworks in detail, the methods used to initialize the interior-point solver are first described.
\subsection*{Initialization methods}
There are four initialization methods: Genetic Algorithm, Random Initialization, Equally Spaced, and manual initialization. 

The Genetic Algorithm used is from the PyMoo library \cite{Pymoo}. A Genetic Algorithm is an evolutionary optimization method inspired by natural selection. It begins with an initial population of candidate solutions, in our case, the design variables \(\mathbf{x}\), which is generated by random initialization. Each individual design variable is evaluated by the nested, ATC, or SOI algorithm. The survival step determines which individuals are retained, typically following the principle of \textit{“survival of the fittest.”} During selection, individuals are chosen as parents based on their fitness. The crossover operator then combines parents to produce offspring. Finally, mutation introduces small random variations in offspring with a certain probability, maintaining diversity and helping the algorithm avoid premature convergence \cite{Pymoo}. The choice for this optimization method is that when a good starting point is found, mutations around that point may lead to an even better solution. Moreover, the random component of the algorithm can occasionally produce entirely new promising starting points. However, there is also a risk that the algorithm converges to a suboptimal starting position.

Next, the Random Initialization function is similar to the first step of the genetic algorithm, as it provides a random selection of design variables \(\mathbf{x}\). The benefit of including both the random and genetic algorithms is that it can be determined whether the evolutionary aspect of the genetic algorithm yields better results than full random.

In addition, we also have an \textit{"equally spaced"} initialization scheme. This initialization places objects and routing nodes in "logical" starting positions by spacing them apart to the corners or halfway points in the design space. This still has a random component; however, it produces a more predictable initial position.

Finally, manual initialization allows a single initialization to be run from a user-specified location.

These initialization methods are used with the algorithms described further in this section.

\subsection*{Nested Framework}
\begin{figure}[t]
\centering
\includegraphics[width=7.5cm]{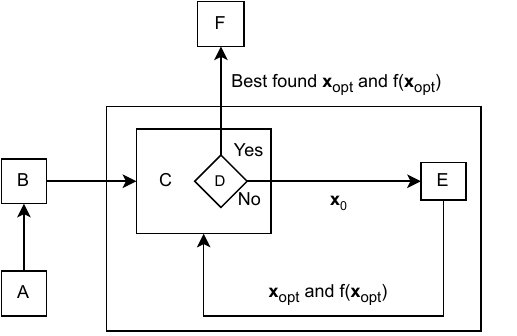}
\caption{Overview of the Nested Approach algorithm. Step A: creation of the MDBD objects. Step B: initialization of object positions and routings. Step C: selection of the initialization method. Step D: evaluation of the exit condition. Step E: optimization using CasADi’s IPOPT solver. Step F: output of the final object placements and routing results.}
\label{fig:Appendix:Nested Aproach}
\end{figure}

The nested framework, shown in Fig.~\ref{fig:Appendix:Nested Aproach}, uses one of four initialization methods to generate an initial position. This initial position, \(\mathbf{x}_0\), is used together with optimization problem~\ref{prob:1} or~\ref{prob:2}, which is solved using CasADi’s IPOPT algorithm~\cite{CasADi}. The resulting objective function value \(f(\mathbf{x}_\mathrm{opt})\) and corresponding design variables \(\mathbf{x}_\mathrm{opt}\) are then returned. These results are stored and used to generate a new initial design variable \(\mathbf{x}_0\)

in the next iteration. The process continues until the exit condition is met, such as reaching a specified number of random initializations or convergence of the genetic algorithm. Once this condition is satisfied, the best objective function value and its corresponding design variables \(\mathbf{x}_\mathrm{opt}\) are selected, from which the final object positions and routing configuration are determined.

\subsection*{Analytical Target Cascading}
Analytical Target Cascading (ATC) is an optimization method designed to solve hierarchical or multilevel optimization problems by decomposing them into smaller, interrelated subproblems~\cite{ATC}. The idea is that this hierarchical decomposition allows complex design problems to be solved more efficiently, supports parallelization, and often improves convergence properties compared to solving a single large-scale problem directly.

\begin{figure}[t]
    \centering
    \includegraphics[width=7.5cm]{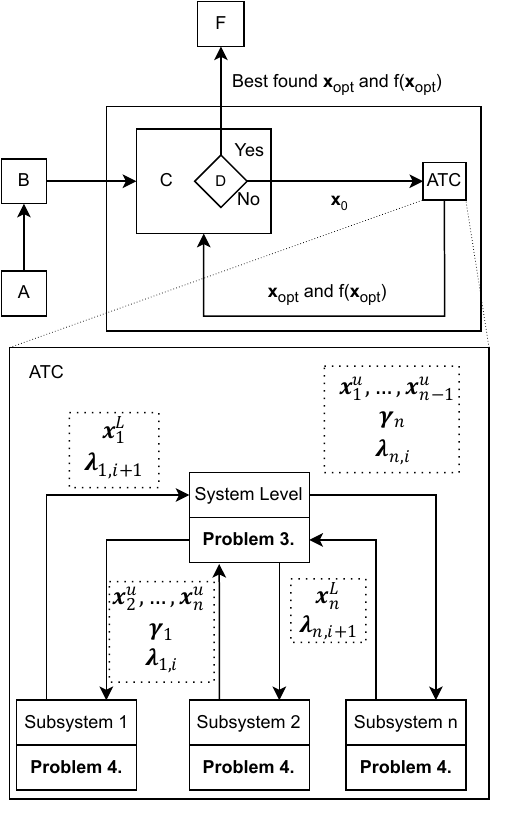}
    \caption{Overview of the Analytical Target Cascading (ATC) framwork. Step A: creation of the MDBD objects. Step B: initialization of object positions and routings. Step C: selection of the initialization method. Step D: evaluation of the exit condition. Step ATC: ATC using CasADi’s IPOPT to iteratively solve Problem \ref{prob:3}. Step F: output of the final object placements and routing results.}
    \label{fig:ATC_Structure}
\end{figure}

In this paper, an initial implementation of ATC is introduced for the placement problem. In this formulation, the routing aspect is omitted, and only the spatial placement of the MDBD objects is considered. As shown in Fig.~\ref{fig:ATC_Structure}, the framework is still very similar to the nested approach, in which a value is provided by an initialization that consists of a system-level component and multiple subsystems, each corresponding to an individual object.

At the system level, the following optimization problem is defined:

\begin{prob}\label{prob:3}
System-Level Optimization Problem
\begin{equation*}
\begin{aligned}
& \min_{\mathbf{x}} && f_\mathrm{v}(\mathbf{x}) + \frac{1}{2}\sum^{n}_{i=1} \pi \|\mathbf{x}^u_i-\mathbf{x}^L_i\|^2 \\
& \text{s.t.} && \mathbf{g}^{\mathrm{obj\mbox{-}obj}}(\mathbf{x}) \leq 0,
\end{aligned}
\end{equation*}
\end{prob}

where \(f_\mathrm{v}\) represents the total volume of the placed objects, \(\mathbf{g}^{\mathrm{obj\mbox{-}obj}}\) are the object--object non-overlap constraints, and \(\pi\) is the scalar coupling parameter (or \emph{trust parameter}) that penalizes deviation between the system-level decisions \(\mathbf{x}^u_i\) (upper-level variables) and the subsystem responses \(\mathbf{x}^L_i\). The design variable \(\mathbf{x}\) is defined by \(\mathbf{x} = [\mathbf{x}^u_1,\dots,\mathbf{x}^u_n]\).

At the subsystem level, each object \(i\) receives its corresponding upper-level decision variables \(\mathbf{x}^u_i\) and solves a local optimization problem of the form:

\begin{prob}\label{prob:4}
Subsystem Optimization Problem
\begin{equation*}
\begin{aligned}
& \min_{\mathbf{x}^L_i} && \zeta\|\mathbf{x}^L_i\|^2 + \boldsymbol{\lambda}_i^\mathrm{T}(\mathbf{x}^L_i - \boldsymbol{\gamma}_i) + \frac{1}{2}\rho\|\mathbf{x}^L_i - \boldsymbol{\gamma}_i\|^2 \\
& \text{s.t.} && \mathbf{g}^{\mathrm{obj\mbox{-}obj}}(\mathbf{x}^L_i,\,\mathbf{x}^u_i) \leq 0,
\end{aligned}
\end{equation*}
\end{prob}

where \(\boldsymbol{\gamma}_i\) denotes the target position communicated from the system level, \(\mathbf{x}^u_i\) are the shared or upper-level variables influencing subsystem \(i\), \(\boldsymbol{\lambda}_i\) is the dual variable enforcing consistency, and \(\rho\) is the penalty weight that determines how strongly each subsystem follows its assigned target. \(\zeta\) is a small stabilization weight which is set to \(10^{-4}\).

The ATC process proceeds iteratively. In each outer iteration, the system broadcasts the updated targets \(\boldsymbol{\gamma}_i\) and upper-level variables \(\mathbf{x}^u_i\) to all subsystems. Each subsystem then solves its local problem and returns its response \(\mathbf{x}^L_i\). The system-level problem is subsequently solved as a \emph{repair step}, restoring feasibility by enforcing the non-overlap constraints and minimizing the total volume.

In less abstract terms, this means that the system minimizes the total volume comprised by the components by setting targets to each of the subsystems. In the subsystem, one MDBD object is moved while all other objects are fixed. By enforcing the constraints, updating the targets, and gradually increasing the coupling and penalty parameters, the system converges to a point where an optimum is reached and \(\mathbf{x} = \mathbf{x}^u = \mathbf{x}^L\). A more detailed explanation on how the targets are updated can be found in Appendix~\ref{Appendix:C}.

\subsection*{Sphere of Influence}
\begin{figure}[t]
    \centering
    \includegraphics[width=7cm]{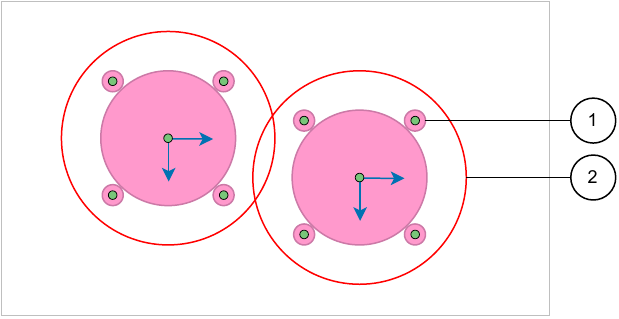}
    \caption{Object Spheres \(b_{i,\mu}\) (1, Light Purple), and Encompassing Sphere (Sphere of Influence) (2, red). Where an overlap in the Encompassing Spheres is seen.}
    \label{fig:Appendix:SOI}
\end{figure}

The final placement approach combines the Geometric Projection (GPM) and Maximal Disjoint Ball Decomposition (MDBD) method to reduce the number of constraint equations. Each object’s MDBD representation is enclosed by a larger GPM sphere. When two GPM spheres are non-overlapping, their corresponding MDBD objects are certainly non-interfering, and no detailed checks are required. However, if two GPM spheres touch or intersect, the objects enter a Sphere of Influence (SOI), where detailed MDBD-level distance constraints are activated, as can be seen in Fig.~\ref{fig:Appendix:SOI}. When the enclosing spheres are disengaged, there are, in this case, two object-object constraint equations. When the enclosing spheres are engaged, there are 25 object-object constraint equations active.

\begin{figure}[t]
    \centering
    \includegraphics[width=8cm]{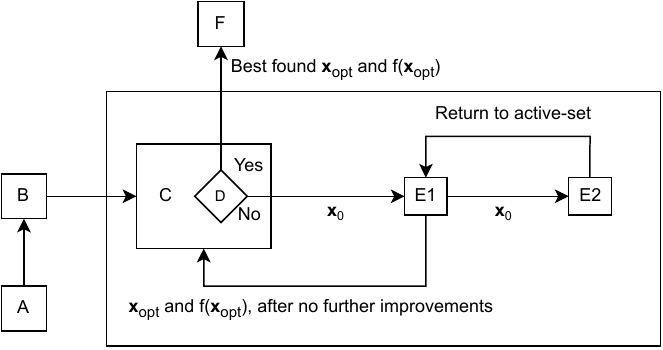}
    \caption{Overview of the Sphere of Influence algorithm. Step A: creation of the MDBD objects, including enclosing sphere. Step B: initialization of object positions. Step C: selection of the initialization method. Step D: evaluation of the exit condition. Step E1: check which constrains must be active based on object position and enclosing spheres interference, Step E2: optimization using CasADi’s IPOPT solver. Step F: output of the final object placements and routing results.}
    \label{fig:Appendix:SOIDiagram}
\end{figure}

In implementation, an active-set strategy dynamically adjusts the number of active constraints during optimization in step E1 (Fig.~\ref{fig:Appendix:SOIDiagram}). Initially, only the constraints between the larger enclosing GPM spheres are considered for all object pairs, resulting in a small and computationally efficient problem. The IPOPT solver in step E2 is then used to refine the current configuration for a limited number of iterations. After this partial solve, the algorithm checks for GPM sphere overlaps. For every overlapping pair, all corresponding MDBD-level sphere–sphere constraints are activated. The optimization is then restarted with this expanded constraint set. This process of solving, checking, and selectively activating constraints repeats until convergence. In essence, the optimizer begins with a coarse collision model and progressively refines it only where necessary, ensuring a fully collision-free placement with the minimal number of active constraints.
\section{Use-Cases}\label{section:use-case}

\subsection{Benchmark Comparison with Prior Work}
A use case defined in a previous study at the University of Connecticut~\cite{Behzadi} is adopted to enable a direct and fair comparison. This allows an objective evaluation between the stochastic gradient descent (SGD) approach of~\cite{Behzadi} and the nested optimization framework developed in this work.

In the original use case, configurations of three, four, and six cubes (each measuring 1.5 × 1.5 × 1.5) were analyzed, each containing 100 MDBD spheres. The cubes were connected by three routing segments that start and end at the center of each MDBD object. The reported quantities include the total volume of the axis-aligned bounding box and the total routing length.

The most relevant setup parameters are summarized in Table~\ref{tab:Apples}. The Genetic Algorithm (GA) is used for initialization. Configurations with 3, 4, and 6 objects are evaluated across various use cases, with three routing segments (two control points) in each. Problem~\ref{prob:1} is solved using the objective functions \(f_1(\mathbf{x})\) and \(f_2(\mathbf{x})\) The objects are modeled as cubes filled with 14, 25, 50, or 100 MDBD spheres, and the optimization algorithm applied is the nested approach. The comparison with exponential routing, denoted by \(f_2(\mathbf{x})\) in Table~\ref{tab:Apples}, is performed only for configurations with 25 spheres per object.

In total, fifteen use cases are evaluated, of which three (those with 100 spheres per object) can be directly compared to~\cite{Behzadi}. Testing multiple MDBD “resolutions” allows assessment of how solution quality varies with the number of spheres per object. Additional setup details are provided in Appendix~\ref{Appendix:E}, and corresponding results are discussed in Section~\ref{section:results}.

\begin{table}[t]
\centering
\small
\caption{Overview of the setup used for the benchmark comparison}
\label{tab:Apples}
\begin{tabularx}{\columnwidth}{l|X|X|X|X}
Number of Spheres &  14 & 25 & 50 & 100\\
\hline
Initialization & \centering Random & \centering Random & \centering Random & \centering Random \tabularnewline
Nr Objects & \centering {3, 4, 6} & \centering {3, 4, 6} & \centering {3, 4, 6} & \centering {3, 4, 6} \tabularnewline
Nr Control Points & \centering 2 & \centering 2 & \centering 2 & \centering 2 \tabularnewline
Problem & \centering 1 & \centering 1 & \centering 1 & \centering 1 \tabularnewline
Objective Function & \(f_1(\mathbf{x})\) & \makecell{\(f_1(\mathbf{x})\),\\ \(f_2(\mathbf{x})\)} & \(f_1(\mathbf{x})\) & \(f_1(\mathbf{x})\) \tabularnewline
Objects & \centering Cubes & \centering Cubes & \centering Cubes & \centering Cubes \tabularnewline
Algorithm & \centering Nested & \centering Nested & \centering Nested & \centering Nested \tabularnewline
\end{tabularx}
\end{table}


\subsection{Novel Benchmark Use-Cases}
As stated previously, the placement and routing problem is a constrained, non-linear, non-convex, and continuous optimization problem. By definition, global optimality cannot be guaranteed or verified. Therefore, a use case is needed where it can be analytically determined whether the desired result is achieved. As shown in the results, this is not possible for the use case introduced in \cite{Behzadi}.  

To address this limitation, several new benchmark cases are introduced to evaluate the performance and robustness of the optimization algorithms. One of the main challenges in this problem is navigating the non-linear loss field. When symmetric use cases are introduced, multiple distinct configurations can yield identical objective values. For example, when placing two identical cubes, there are 144 configurations that lead to the same minimal volume, indicating a high degree of symmetry.  

For routing, the resulting configuration must also be logically interpretable and analytically optimal. Additionally, the objectives should not be conflicting: when the volume is minimal, the routing distance should be minimal as well. This is particularly important since IPOPT from CasADi does not directly handle multi-objective optimization.

Based on these considerations, three novel benchmark use cases are defined: one with high symmetry, one with reduced symmetry, and one with a unique configuration.

\subsubsection{Cuboids}
\begin{figure}[t]
    \centering
    \includegraphics[width=9cm]{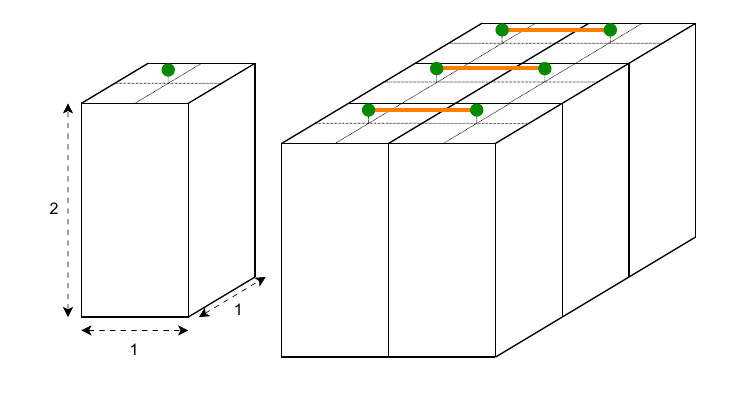}
    \caption{Cuboid Schematic: the dimensions 1x1x2 are shown in the single cuboid on the left; the port of the cuboid is shown in green. Note that there are two ports; only the bottom one is not visible. On the right-hand side of the figure, an analytical optimal configuration of 6 cuboids with routing is shown.}
    \label{fig:Cuboid}
\end{figure}
For the high-level of symmetry case, a rectangular 1x1x2 shape is used with two ports for routing. These ports are connected at the long end. Fig.~\ref{fig:Cuboid} displays the geometry of one object on the left-hand side and 6 objects on the right-hand side. The port connectivity for two objects is as follows \(\mathrm{connections} = [(\varphi_{1,1},\varphi_{2,1}), (\varphi_{1,2},\varphi_{2,2})]\) which means that port one of object one is connected to port one of object two, and port two of object one is connected to port two of object two. This scales for four and six objects, as shown in Fig.\ref{fig:Cuboid}. This use case provides multiple analytical optimal configurations. For two objects, the volume should be four and the routing length two, for four objects eight and four, and for six objects twelve and six. Additionally, the optimal positions should be easily distinguishable from sub-optimal ones.

In terms of symmetry, for two objects, there are sixteen possible configurations that are equally minimal in both volume and routing. For four objects, 2048 configurations, and with six objects even more configurations. These configurations arise from the combinatorial symmetry of the cuboid geometry and its port connectivity. Each object can be mirrored along its longitudinal or lateral axis and rotated by 90°, 180°, or 270° about the global axes without changing the total occupied volume or the routing distance. Because all objects are identical, every mirrored or rotated arrangement that preserves the port alignment produces an equivalent optimum. 

\subsubsection{L-shapes}
\begin{figure}[t]
    \centering
    \includegraphics[width=9cm]{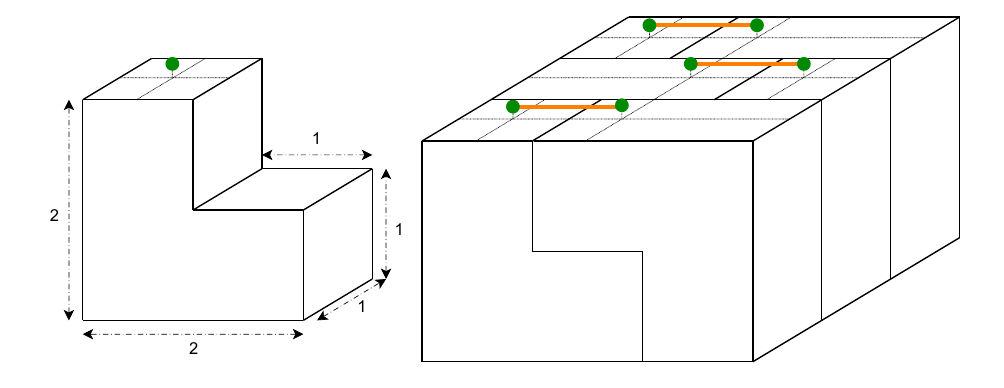}
    \caption{L-Shape Schematic; with on the left a single L-shape and on the right an analytical optimum of six L-shaped objects is displayed.}
    \label{fig:L_Shape}
\end{figure}
To reduce the number of possible configurations, we can also use an L-shape with routing, as shown in Fig.~\ref{fig:L_Shape}. The port connectivity for two objects would be \(\mathrm{connections} = [(\varphi_{1,1},\varphi_{2,2}), (\varphi_{1,2},\varphi_{2,1})]\).

The use case provides an analytical optimum: for two objects, the minimal volume is six with a minimal routing distance of two; for four objects, the minimal volume is twelve with a routing distance of four; and for six objects, the minimal volume is eighteen with a routing distance of six. An example of the configuration is shown on the right-hand side of Fig.~\ref{fig:L_Shape}. 

For two L-shaped objects, the configuration is unique. For four objects, there are 192 possible configurations.

\subsubsection{Unique}
\begin{figure}[t]
    \centering
    \includegraphics[width=9cm]{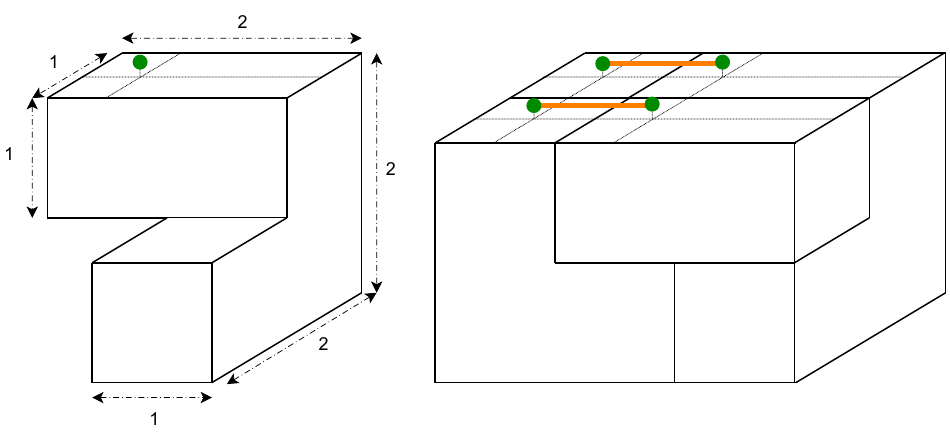}
    \caption{Unique benchmark schematic. The left shows the double L-shape geometry. On the right, this double L-shape is combined with two L-shapes and one cuboid, forming a configuration with four routing paths. The L-shapes are also connected to the double L-shape at the bottom.}
    \label{fig:Unique}
\end{figure}
The final benchmark represents a unique configuration, meaning that only one solution exists. As discussed in the previous section, the two-object L-shape already yields a unique arrangement. Here, the objective is to construct a similarly unique use case for four objects to enable direct comparison with the symmetric cuboid and L-shape benchmarks. A six-object version is omitted, as the focus is on examining how symmetry influences the number of analytical optima identified by the initialization algorithms.

As shown in Fig.~\ref{fig:Unique}, the configuration consists of a new base geometry on the left, combined with two L-shapes and one cuboid featuring distinct port positions. The total volume of the configuration is twelve, and the routing length is four, with two routing connections located on the underside (not visible in Fig.~\ref{fig:Unique}). The corresponding MDBD representations of these objects are provided in Appendix~\ref{Appendix:F}.

\subsection{Test Plan: What to Test and Why}
The aim of this study is to evaluate whether the proposed nested hybrid framework can solve spatial placement and routing \emph{accurately} and \emph{efficiently}, and to understand when alternative formulations (constraints, decomposition, routing objectives) help or hurt. Following the gaps identified in the introduction, we structure the tests around six questions:

\begin{itemize}
    \item Accuracy vs.\ ground truth (Cuboid benchmark): How close are converged solutions to analytical optima?
    \item Robustness to initialization: How sensitive are outcomes to Equally Spaced (ES), Genetic Algorithm (GA), and Random starts?
    \item Symmetry: How does problem symmetry (Cuboid, L-shape, Unique) affect the share of near-optimal solutions?
    \item Framework choice: How do Nested, ATC, and SOI compare in objective value and runtime on identical tasks?
    \item Constraint and objective formulations: What is the trade-off between Absolute vs.\ Soft-Sum constraints, and between quadratic vs.\ exponential routing objectives?
    \item Scalability and resolution: How do iteration-level times scale with numbers of objects, spheres (MDBD resolution), and routing control points; and does warm-starting from lower resolution help?
\end{itemize}

Across tests, solver tolerances, penalty weights, bounds, and stopping criteria are held fixed. When comparing frameworks (Nested/ATC/SOI) or formulations (Absolute/Soft-Sum; quadratic/exponential routing), only the factor under study is varied.

Table~\ref{tab:TestInit} lists the experimental factors and levels used throughout the study. Not all combinations are exercised; we select minimal, targeted subsets per research question (full schedule in Appendix~\ref{Appendix:G}).
\begin{table}[t]
\centering
\small
\caption{Experimental factors and levels used across tests}
\label{tab:TestInit}
\begin{tabularx}{\columnwidth}{l|X|X|X}
Parameter & Setting A & Setting B & Setting C \\
\hline
Initialization & \centering Random & \centering GA & \centering ES \tabularnewline
Nr Objects & \centering 2 & \centering 4 & \centering 6 \tabularnewline
Nr Spheres & \centering 20 & \centering 40 & \centering 60 \tabularnewline
Nr Control Points & \centering no routing & \centering 2 & \centering 4 \tabularnewline
Problem & \centering 1 & \centering 2 & \centering -- \tabularnewline
Objective Function & \centering \(f_3(\mathbf{x})\) & \centering \(f_4(\mathbf{x})\) & \centering -- \tabularnewline
Objects & \centering Cuboids & \centering L-shape & \centering Unique \tabularnewline
Algorithm & \centering Nested & \centering ATC & \centering SOI \tabularnewline
\end{tabularx}
\end{table}
\begin{table}[t]
\centering
\small
\caption{Warm-start schedule from 20-sphere solutions (Cuboid).}
\label{tab:ByHand}
\begin{tabularx}{\columnwidth}{l|X|X|X}
Parameter & Setting A & Setting B & Setting C\\
\hline
Initial Position & \centering best \(\mathbf{x}_0\) & \centering \(\mathbf{x}_\mathrm{opt}\) & \centering -- \tabularnewline
Nr Objects & \centering 2 & \centering 4 & \centering 6 \tabularnewline
Nr Spheres & \centering 30 & \centering 40 & \centering -- \tabularnewline
Nr Control Points & \centering no routing & \centering 2 & \centering -- \tabularnewline
Problem & \centering 1 & \centering -- & \centering -- \tabularnewline
Objective Function & \centering \(f_3(\mathbf{x})\) & \centering -- & \centering -- \tabularnewline
Objects & \centering Cuboids & \centering -- & \centering -- \tabularnewline
Algorithm & \centering Nested & \centering -- & \centering -- \tabularnewline
\end{tabularx}
\end{table}

To study resolution scaling, higher-resolution runs are warm-started from previously obtained lower-resolution results. Two initializations are used: the best-found initial point $\mathbf{x}_0$ and the best converged solution $\mathbf{x}_{\mathrm{opt}}$ at 20 spheres. Both $\mathbf{x}_0$ and $\mathbf{x}_{\mathrm{opt}}$ contain the same design variables—namely the object frame position $(\mathbf{x}_{A_i})$ and the routing control point positions $(\mathbf{c}_{L,k})$. These are directly reused as initial guesses for higher-resolution targets with 30–40 spheres, with and without routing (Table~\ref{tab:ByHand}), where the additional spheres are generated around the same transformed object frames to preserve the spatial configuration.


Exercising Table~\ref{tab:TestInit} combinatorial would yield 2{,}916 cases and is out of scope. We therefore test only the contrasts needed to answer the six questions above, matching the comparisons shown in the Results.

\section{Results}\label{section:results}
This section reports: \textit{1)} a comparison of the three frameworks; Nested, Analytical Target Cascading (ATC), and Sphere of Influence (SOI); \textit{2)} an apples-to-apples comparison against \cite{Behzadi}; \textit{3)} accuracy on the Cuboid benchmark relative to ground truth; \textit{4)} initialization performance (share of runs within 0.5\% of the best found objective); \textit{5)} robustness across Cuboid, L-shape, and Unique use cases; \textit{6)} iteration-level runtime scaling in CasADi IPOPT; and \textit{7)} two strategies to reduce computation time (Soft-Sum constraints and warm-starting from a previous best).

All experiments ran on a 64-bit Windows laptop with an Intel\textsuperscript{\textregistered} Core\textsuperscript{TM} i7-7700HQ @ 2.80,GHz, 8,GB RAM, a 238,GB Samsung SSD, and an NVIDIA Quadro M1200 (4,GB) plus integrated Intel\textsuperscript{\textregistered} HD Graphics 630. The framework is implemented in Python~3.11.9~\cite{Python}.

\subsubsection{Comparing Frameworks}

\begin{table}[t]
\centering
\small
\caption{Results for 20 spheres, no routing, Cuboid case, with 100 random initializations}
\label{tab:20SpheresNoRouting}
\begin{tabularx}{\columnwidth}{l|X|X|X}
Method & Objects & Total Time [s] & Best Result \\
\hline
Nested & \centering 2 & \centering 70.95 & \centering 3.92 \tabularnewline
Nested & \centering 4 & \centering 749.16 & \centering 7.87 \tabularnewline
ATC & \centering 2 & \centering 909.92 & \centering 3.92
 \tabularnewline
ATC & \centering 4 & \centering 29455.90 & \centering 7.87
 \tabularnewline
SOI & \centering 2 & \centering 75.04 & \centering 3.92
 \tabularnewline
SOI & \centering 4 & \centering 1056.87 & \centering 7.86 \tabularnewline
\end{tabularx}
\end{table}

Table~\ref{tab:20SpheresNoRouting} compares the Nested, ATC, and SOI frameworks for the Cuboid case with 20 spheres, no routing, and 100 random initializations. The ``Best Result'' column shows the lowest objective value obtained across all runs.

All three methods converge to nearly identical minima for both two and four objects, confirming that the problem formulation is consistent across frameworks. However, the computational cost differs significantly. The Nested framework is the fastest, followed by SOI, while ATC is the slowest by over an order of magnitude.

The SOI’s slower performance is mainly due to CasADi’s symbolic graph being rebuilt each time the active constraint set changes. For ATC, both system- and subsystem-level constraints (see \textbf{Problems~\ref{prob:3}} and \textbf{\ref{prob:4}}), are evaluated multiple times per iteration, which scales poorly with the number of subsystems. Although ATC has the potential to be faster if the design space and constraint equations are efficiently partitioned, the results in this study indicate that this advantage was not realized here. 

\subsubsection{Comparative Use Case}
\begin{table}[t]
\centering
\small
\caption{Comparison between the Stochastic Gradient Descent (SGD) method used in \cite{Behzadi} and the results generated in this study. All objects contain 100 spheres. The configurations are visualized in Fig.~\ref{fig:3objects}.}
\label{tab:ReferenceSolution}
\begin{tabularx}{\columnwidth}{l|X|X|X|X|X}
Initialization & Objects & Volume & Routing Length & Sum & Difference \% to \cite{Behzadi} \\
\hline
SGD~\cite{Behzadi} & \centering 3 & \centering 12.24 & \centering 7.06 & \centering 19.30 & \centering -- \tabularnewline
Random & \centering 3 & \centering 10.60 & \centering 6.54 & \centering 17.14 & \centering --11.19 \tabularnewline
SGD~\cite{Behzadi} & \centering 4 & \centering 16.40 & \centering 6.16 & \centering 22.56 & \centering -- \tabularnewline
Random & \centering 4 & \centering 13.60 & \centering 6.17 & \centering 19.77 & \centering --12.37 \tabularnewline
SGD~\cite{Behzadi} & \centering 6 & \centering 25.66 & \centering 13.31 & \centering 38.97 & \centering -- \tabularnewline
Manual & \centering 6 & \centering 24.11 & \centering 17.95 & \centering 42.06 & \centering 7.93
\end{tabularx}
\end{table}

\begin{table}[t]
\centering
\small
\caption{Random initialization using the quadratic routing objective function. The difference is compared to \cite{Behzadi}.}
\label{tab:25Q}
\begin{tabularx}{\columnwidth}{l|X|X|X|X|X}
Spheres & Objects & Volume & Routing Length & Sum & Difference\% to \cite{Behzadi} \\
\hline
\centering 25 & \centering 3 & \centering 10.80 & \centering 6.70 & \centering 17.50 & \centering -9.33 \tabularnewline
\centering 25 & \centering 4 & \centering 13.63 & \centering 6.46 & \centering 20.09 & \centering -10.95 \tabularnewline
\centering 25 & \centering 6 & \centering 21.05 & \centering 14.91 & \centering 35.96 & \centering -7.72 \tabularnewline
\centering 50 & \centering 6 & \centering 21.13 & \centering 16.68 & \centering 37.81 & \centering -2.98
\end{tabularx}
\end{table}

\begin{table}[t]
\centering
\small
\caption{Comparison of exponential to quadratic routing objective functions for 25-sphere initialization. The difference is compared to the quadratic routing objective results from Table~\ref{tab:25Q}}
\label{tab:25E}
\begin{tabularx}{\columnwidth}{l|X|X|X|X|X}
Spheres & Objects & Volume & Routing Length & Sum & Difference\% to Table~\ref{tab:25Q} \\
\hline
Random & \centering 3 & \centering 13.75 & \centering 5.09 & \centering 18.85 & \centering 7.69  \tabularnewline
Random & \centering 4 & \centering 15.74 & \centering 10.55 & \centering 26.29 & \centering 30.88 \tabularnewline
Random & \centering 6 & \centering 52.59 & \centering 25.70 & \centering 78.29 & \centering 117.72
\end{tabularx}
\end{table}

\begin{figure}[t]
    \centering
    \includegraphics[width=9.1cm]{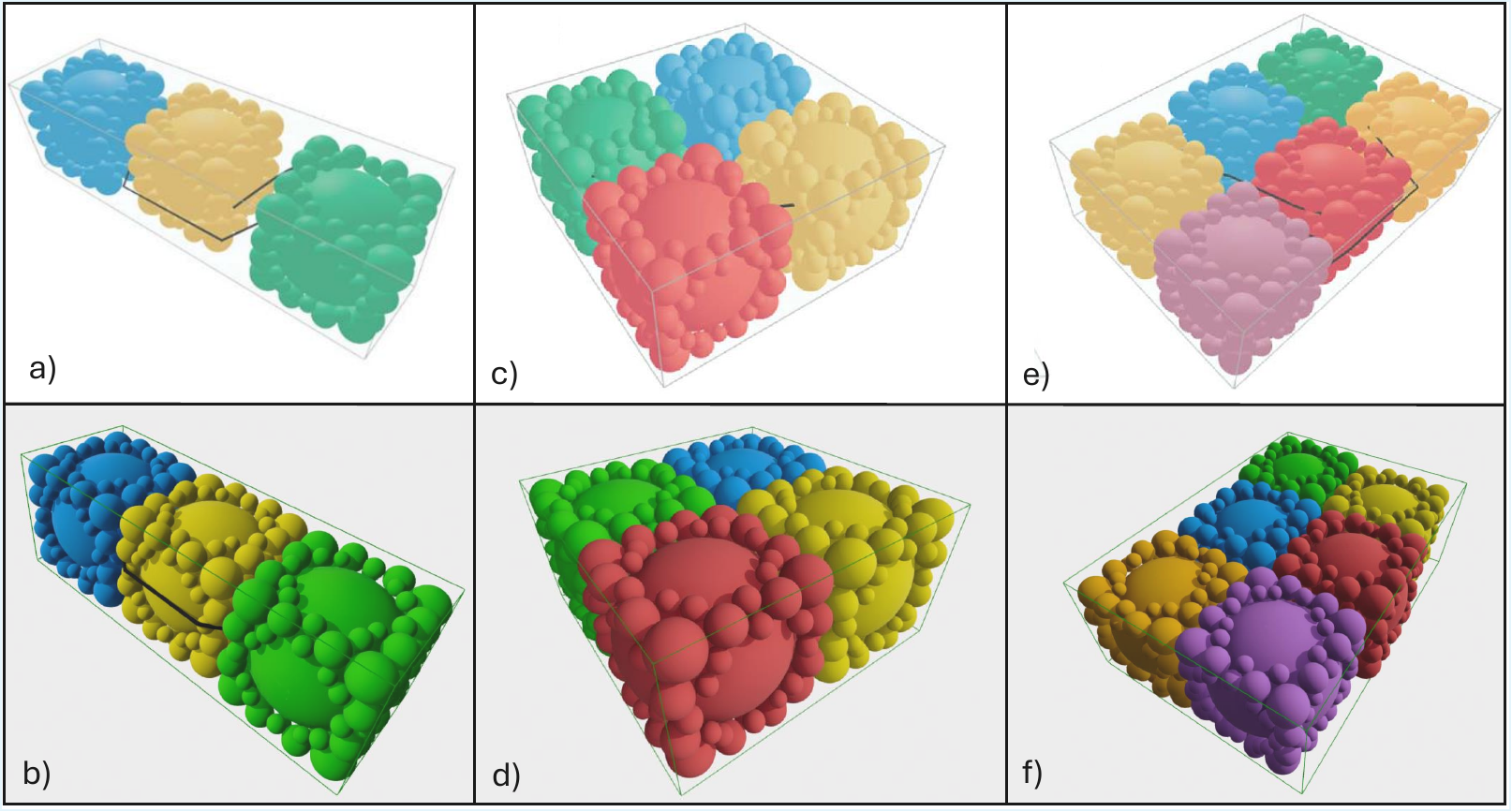}
    \caption{Apples-to-apples comparison for configurations with 100 spheres. Results (a), (c), and (e) are reproduced from \cite{Behzadi}, while (b), (d), and (f) are generated with the optimization framework proposed in this study.}
    \label{fig:3objects}
\end{figure}

Table~\ref{tab:ReferenceSolution} and Fig.~\ref{fig:3objects} compare our nested framework to the SGD baseline of \cite{Behzadi}. With 100 spheres, our method improves the combined placement–routing objective by more than 10\% for three and four objects. For six objects, improvements are obtained at 50 spheres or fewer; with 100 spheres, solver convergence requires manual initialization, limiting the benefit of multiple initializations. The reason why the presented nested hybrid algorithm outperforms the Stochastic Gradient Descent could be a combination of its exploitation using more curvature information (via the Hessian) and its robust handling of nonlinear constraints. Unlike SGD, the interior-point optimization step leverages second-order derivatives and barrier functions to ensure smooth convergence within feasible regions, while maintaining strict constraint satisfaction throughout the process.

Tables~\ref{tab:25Q}–\ref{tab:25E} further show that, in this setup, a quadratic routing objective outperforms an exponential one. The quadratic term provides a smoother and better-conditioned optimization landscape, allowing the solver to converge more reliably and consistently across initializations. In contrast, the exponential formulation amplifies small variations in routing length, leading to ill-conditioning and dominance over other objective terms. As a result, the quadratic objective achieves lower final costs and improved stability.

\subsubsection{Precision Against Ground Truth}
\begin{figure}[t]
    \centering
    \includegraphics[width=9cm]{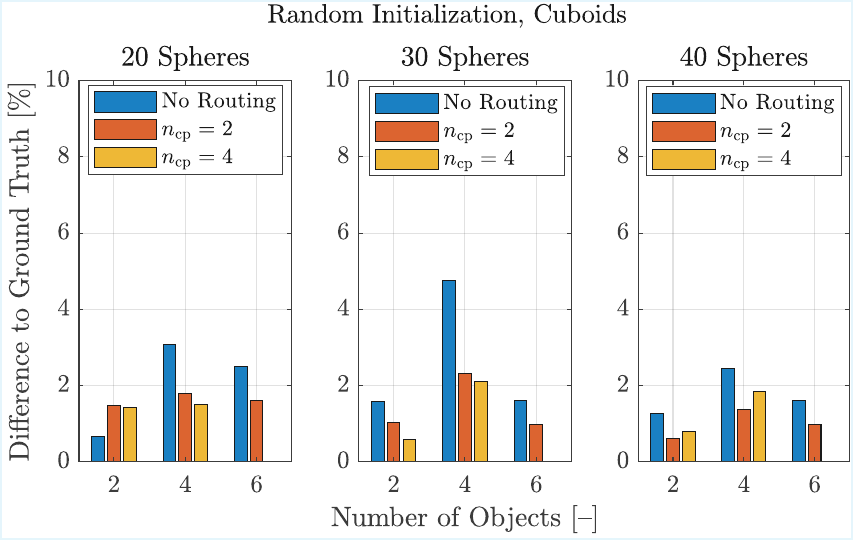}
    \caption{The figure displays the difference between the converged solution and the ground truth of the Cuboid use-case}
    \label{fig:Results5}
\end{figure}
Using the new Cuboid benchmark with known optima, we quantify accuracy as the percent deviation of the converged solution from the ground truth volume and routing length. For example, two cuboids without routing have minimal volume 4; with 20 spheres and random initialization we obtain 4.03 (0.75\% above optimum). As summarized in Fig.~\ref{fig:Results5}, most cases fall within roughly 0.6–2.0\% of ground truth, with a few outliers exceeding 2.0\%. Zero bars indicate non-convergence.

\subsubsection{Initialization}
\begin{table}[t]
\centering
\small
\caption{Number of solutions within 0.5\% of the best found result for different initializations. The cuboid use case is used with two objects and no routing connections}
\label{tab:ESGaRA}
\begin{tabular}{l|l|l}
\hline
Initialization & Number of Spheres & Solutions within $0.5\%$ of $f_\mathrm{opt}$ \\
\hline
ES      & 20 & 76\% \\
GA      & 20 & 54.44\% \\
Random  & 20 & 48\% \\
ES      & 30 & 65.33\% \\
GA      & 30 & 61.67\% \\
Random  & 30 & 50\% \\
ES      & 40 & 81.33\% \\
GA      & 40 & 64.44\% \\
Random  & 40 & 67\% \\
\end{tabular}
\end{table}

\begin{figure}[t]
    \centering
    \includegraphics[width=9cm]{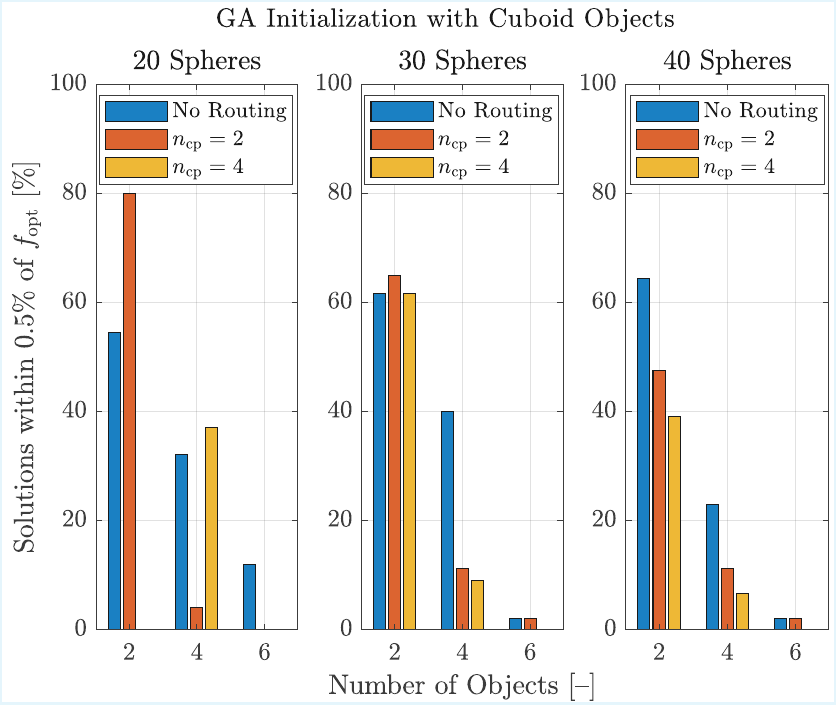}
    \caption{Results of initialization with Genetic Algorithm using the nested optimization framework. The number of objects, spheres, and routing connections is varied. The cuboid use case is used}
    \label{fig:Results3}
\end{figure}

\begin{figure}[t]
    \centering
    \includegraphics[width=9cm]{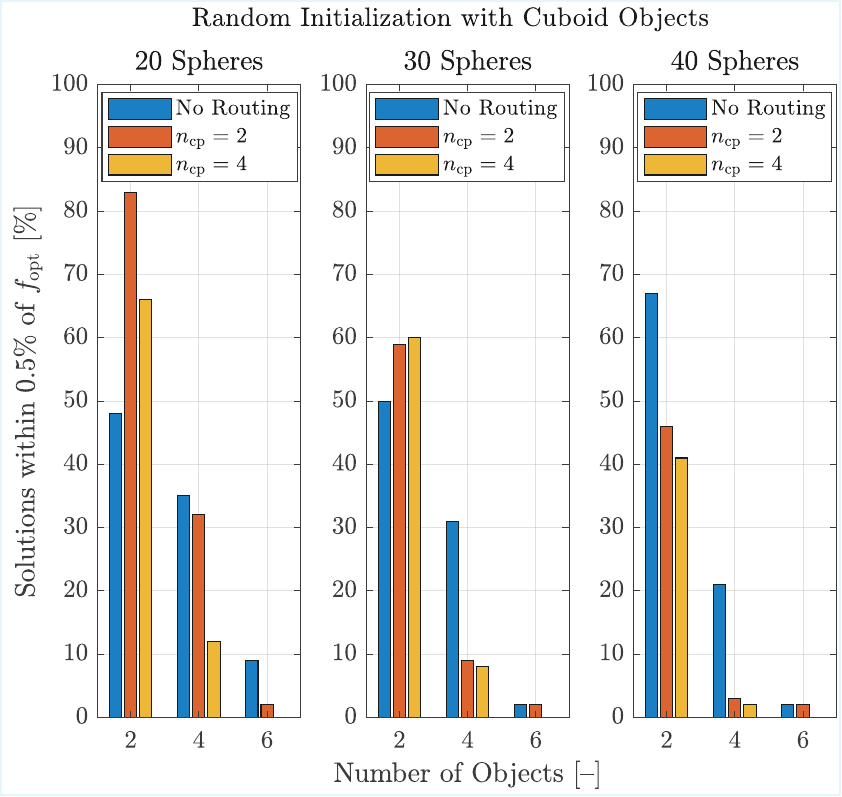}
    \caption{Results with randomized initialization using the nested optimization framework. The number of objects, spheres, and routing connections is varied. The cuboid use case is used}
    \label{fig:Results6}
\end{figure}
Table~\ref{tab:ESGaRA} reports the share of runs within 0.5\% of $f_\mathrm{opt}$ for Equally Spaced (ES), Genetic Algorithm (GA), and Random initialization (two objects, no routing). ES performs best overall, followed by GA, then Random. However, ES is unreliable when routing is present; therefore, subsequent routing experiments use GA or Random only. Figs.~\ref{fig:Results3} and \ref{fig:Results6} show that the fraction of near-optimal solutions decreases as the number of objects increases; for six objects with four control points, no solutions are found.

\subsubsection{Various Use Cases}

\begin{table}[t]
\centering
\small
\caption{Number of solutions within 0.5\% of the best found result for different symmetry cases. Case: 4 objects, 20 spheres, 2 routing points. Initialized with a genetic algorithm.}
\label{tab:Symmetry}
\begin{tabular}{l|l}
\hline
Use Case & Solutions within $0.5\%$ of $f_\mathrm{opt}$ \\
\hline
Cuboid & 4\% \\
L-Shape & 1\% \\
Unique & 0\% \\
\end{tabular}
\end{table}
\begin{figure}[t]
    \centering
    \includegraphics[width=9cm]{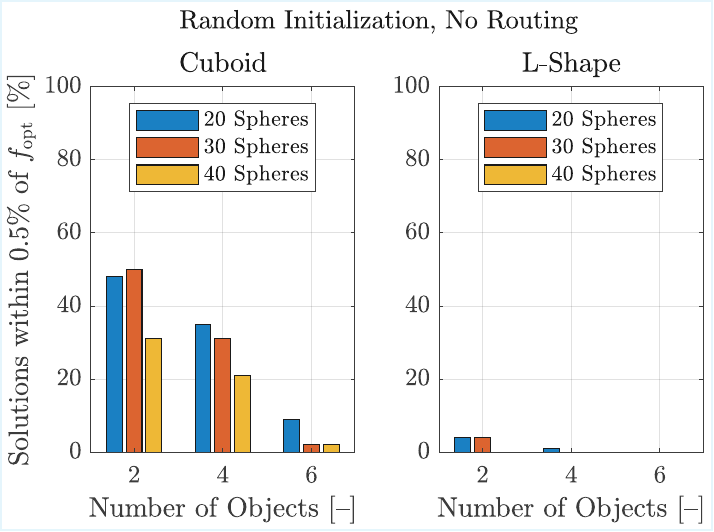}
    \caption{In this figure the number of solutions within 0.5\% of the best found result are compared between the Cuboid use-case and the L-Shape use-case}
    \label{fig:Results2}
\end{figure}




We test Cuboid, L-shape, and Unique configurations. With four objects, two control points, and 20 spheres (GA initialization), Table~\ref{tab:Symmetry} shows that Cuboid and L-shape reach optima in a small share of runs, whereas the Unique case does not converge to the expected target (see Section~\ref{section:use-case}). For the L-shape without routing (Fig.~\ref{fig:Results2}), four objects and 20 spheres yield optimal or near-optimal solutions in about 2\% of initializations; increasing objects or spheres eliminates successful runs. These results indicate a practical complexity limit for the present formulation and solver settings.

\subsubsection{Time-complexity}
\begin{figure}[t]
    \centering
    \includegraphics[width=9cm]{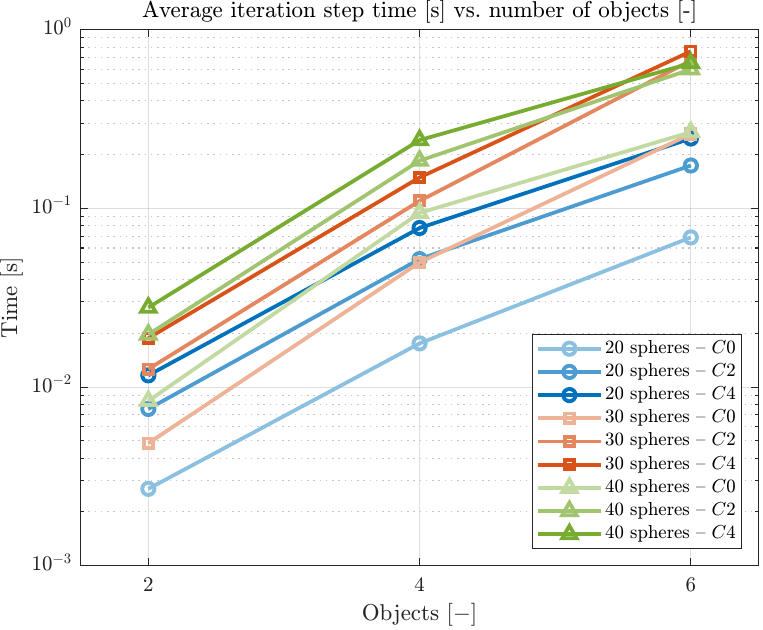}
    \caption{CasADi inter point optimization iteration step time, compared with number of objects, spheres, and control points}
    \label{fig:Time}
\end{figure}

Fig.~\ref{fig:Time} shows the average iteration time of CasADi’s IPOPT solver as a function of the number of objects, spheres, and routing control points. The iteration time reflects the computational complexity of a single interior-point step, independent of initialization or convergence behavior. Although the total runtime to reach an optimum also depends on the number of iterations— which can vary significantly with initialization and problem scaling— the per-iteration time provides a fairer measure of intrinsic problem complexity. As seen in Fig.~\ref{fig:Time}, the cost per iteration increases almost exponentially with the number of objects and spheres, whereas increasing the number of routing control points from two to four has a comparatively minor effect.

\subsubsection{Attempts to improve solving speed}
\begin{figure}[t]
    \centering
    \includegraphics[width=9cm]{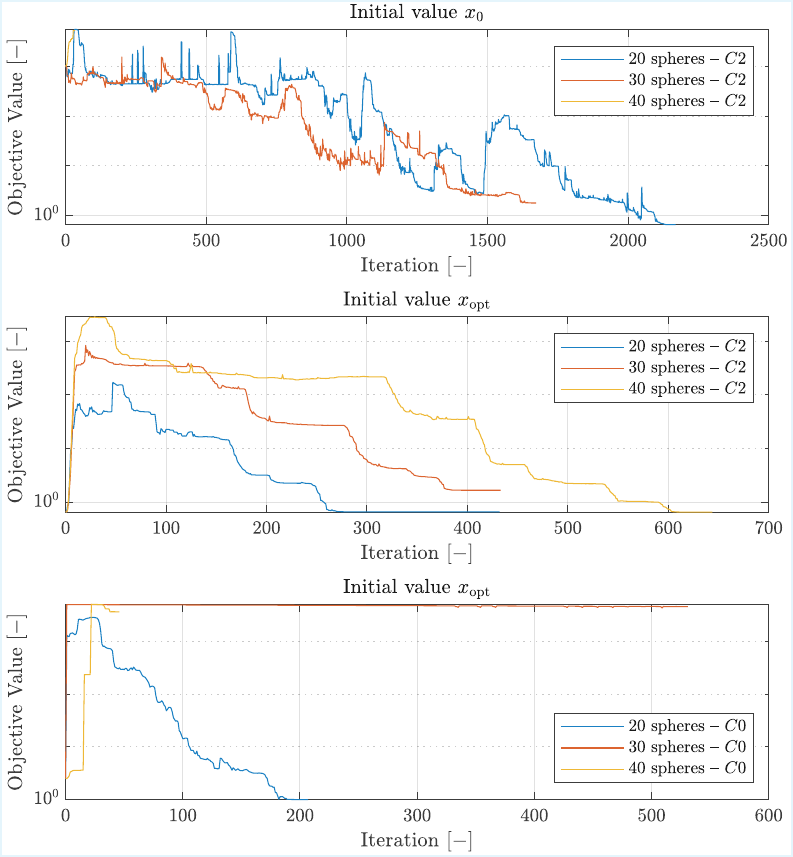}
    \caption{Different Initializations from previous best $x_0$ and $x_\mathrm{opt}$}
    \label{fig:PreviousBest}
\end{figure}
\begin{table}[t]
\centering
\small
\caption{Comparison between SoftSum and Absolute constraints for the Cuboid use case with 20 spheres}
\label{tab:SoftSumVsAbsolute}
\begin{tabularx}{\columnwidth}{l|X|X|X|X}
\textbf{Constraint Type} & \textbf{Objects} & \textbf{CasADi Iter Time [s]} & \textbf{Solution} \\
\hline
Absolute & \centering 2 & \centering 0.007933154 & \centering Good \tabularnewline
SoftSum & \centering 2 & \centering 0.002303885 & \centering Good, small interference \tabularnewline
Absolute & \centering 4 & \centering 0.054830765 & \centering Good \tabularnewline
SoftSum & \centering 4 & \centering 0.008112342 & \centering Not found \tabularnewline
Absolute & \centering 6 & \centering 0.169434113 & \centering Not found \tabularnewline
SoftSum & \centering 6 & \centering 0.008112342 & \centering Good, small interference \tabularnewline
\end{tabularx}
\end{table}
We evaluate two tactics. First, we warm-start higher-resolution problems from the best solution at lower resolution (e.g., solve 20-sphere instances with multiple starts, then use the best $x_0$ to launch 30-/40-sphere runs). As shown in Fig.~\ref{fig:PreviousBest}, outcomes vary: some 40-sphere runs diverge to infeasibility, others recover the 20-sphere optimum; without routing, both 30- and 40-sphere runs from $x_\mathrm{opt}$ can become infeasible.

Second, we replace Absolute constraints (Problem~\ref{prob:1}) with Soft-Sum constraints (Problem~\ref{prob:2}), which reduce the constraint count to three regardless of problem size. Table~\ref{tab:SoftSumVsAbsolute} shows substantial iteration-time reductions as objects increase. The trade-off is weaker separation—occasionally yielding small interferences—and, in some settings, failure to find a feasible solution.

\subsection*{Discussion}
A few remarks are in order. Each test in this study was executed once, meaning that the initialization results do not constitute a rigorous sensitivity analysis. The aim, however, was to obtain a first-order understanding of how well the proposed framework generally finds feasible and near-optimal solutions. Running single instances per condition offers a straightforward way to observe solver behavior as problem complexity increases with the number of spheres, routing connections, and objects.



\section{Conclusion}\label{section:conclusion}
This study introduced a nested hybrid optimization framework for spatial placement and routing within the SPI2 context, combining stochastic exploration with deterministic refinement through IPOPT and CasADi. The framework was validated using analytically verifiable benchmarks, enabling a direct quantitative assessment of optimizer accuracy—an aspect previously missing from the SPI2 literature.

Across benchmarks, the nested approach achieved near-optimal solutions within 0.6–2\% off the analytical ground truth for most cases, where 0\% difference is the best result. It outperformed the stochastic gradient descent baseline from~\cite{Behzadi} by over 10\% for 3 and 4 objects. For six objects and 100 spheres, the framework performs worse. The test had to be manually initialized and therefore does not take advantage of the stochastic initialization. This is most likely due to normalization and not an inherent limitation of the framework.

Comparative evaluations with Analytical Target Cascading (ATC) and Sphere of Influence (SOI) showed that all three frameworks can reach the same minima but with substantial differences in computational cost. The nested approach proved the most efficient for smaller systems, while SOI may gain advantages as the number of objects increases. Initialization analysis further indicated that higher symmetry increases the number of near-optimal results, whereas lower symmetry reduces the number of found solutions.

Runtime scaling analysis showed that iteration time grows nearly exponentially with the number of objects and spheres, underscoring the importance of decomposition and constraint-reduction methods. Replacing pairwise Absolute constraints with aggregated Soft-Sum formulations reduced iteration time by up to an order of magnitude; however, it reduced solution accuracy significantly.

Taken together, the results confirm that a hybrid optimization approach can effectively handle nonlinear and non-convex spatial placement–routing problems, offering improvements in accuracy over previous MDBD-based SPI2 placement research. Moreover, the proposed verifiable benchmarks and structured testing protocol provide a foundation for future comparison across SPI2 frameworks.
\section{Future Work and Recommendations}\label{sec:Future}
Future research can proceed along several directions. Extending functionality and physics, the current framework can be expanded to include additional functional and physical modeling capabilities. Functionality-wise, adding angular or kinematic constraints to routing paths would enable the representation of rotating shafts and axles. On the physical side, integrating effects such as temperature, electrical resistance, center of gravity, and moment of inertia would allow optimization under coupled multiphysics conditions. In addition, adding a custom boundary box, which is not a rectangle like the axis-aligned boundary box, improves its usefulness in real engineering scenarios, as engineering almost never happens in a square box. All aligning the framework more closely with real engineering applications.

Improving computational efficiency: Replacing rotation matrices with quaternion-based orientation handling would prevent gimbal lock and improve numerical stability. Furthermore, the framework should be parallelized to leverage modern multi-core hardware: running multiple stochastic initializations (e.g., Random or GA) in parallel across 32–64 cores could drastically increase the number of independent starts and improve the likelihood of finding analytical solutions for more complex geometries, such as the Unique use case.

This also brings us to the warm starting of solutions from previous best solutions with less sphere resolution. Right now, it does not yet scale. However, with changes to the IPOPT Casadi settings, the situation might improve. 

Refining stochastic initialization strategies; Currently, each initialization is directly followed by a full IPOPT run. A more efficient two-stage process could first screen a large set of low-cost random initializations (e.g., 1000 trials) using a simplified surrogate or gradient-free evaluation, then select promising candidates for refinement with IPOPT. Such staged or adaptive initialization strategies may better balance exploration and convergence. This is just one example; there are many options for trying different initialization strategies.

Multi-objective optimization: A key limitation of IPOPT is its single-objective formulation. Incorporating a gradient-based solver capable of handling multi-objective optimization would enable simultaneous evaluation of conflicting criteria, such as minimizing volume and routing length in scenarios where minimizing one is maximizing the other. Pareto-based formulations would provide engineers with trade-off insights and improve decision-making.

Advancing decomposition methods; While Analytical Target Cascading (ATC) performed worst among the tested frameworks, its underlying principle of spatial decomposition remains promising. Alternative decomposition schemes or adaptive partitioning strategies may yield much greater scalability if subproblems are defined and coordinated more effectively. Similarly, the Sphere of Influence (SOI) approach, though not yet the fastest, points toward a promising direction for localized solving within decomposed regions.

Beyond traditional optimization, the spatial placement and routing problem lends itself naturally to learning-based methods. Reinforcement learning or other unsupervised machine learning approaches could autonomously explore the design space, treating placement and routing as decision problems with well-defined rewards. Such integration could enable rapid generalization across geometries and potentially discover design strategies beyond those accessible through conventional optimization.

\section*{Acknowledgements}
This report has been partially prepared with the assistance of generative AI tools (OpenAI ChatGPT). These tools were used to improve the structure, clarity, and consistency of the writing, as well as to format the LaTeX document. All ideas, analysis, and technical content originate from the authors, and the final text has been reviewed and verified for accuracy and originality.
\FloatBarrier

\cleardoublepage
\appendices
\section{}\label{Appendix:A}
There are no standard programs yet to create Maximal Disjoint Ball Decomposition (MDBD) objects. The master thesis \cite{Vu} describes a method to create a MDBD Object, however, this script is not publicly available. In this Appendix we describe how a new method and algorithm is constructed to be able to transform meshes into MDBD objects with a Brute-Force Methodology. 

This Appendix is setup as follows, first, the methods of the Brute-Force program is explained, second, which test cases are used to validate the functionality of the algorithm, third the results. And finally, a brief conclusion and discussion.

\subsection{Methods for MDBD}
The main function of the MDBD script is to fill a 3D object defined by a mesh with disjoint spheres. Consider a watertight mesh \(\Phi_{\mathrm{mesh}}\), this file is placed in a workspace \(\Omega \subset \mathbb{R}^3\) which is a rectangular grid domain with nodes \(\{x_{ijk}=(x_i,y_j,z_k)\}_{i,j,k=1}^{N_\mathrm{r}}\). 

Now, nodes can be assigned a value based on where they are compared to the mesh. When a point is outside of the mesh and therefore is infeasible the value of the node becomes \(-\infty\). When the position is possible the node obtains the value of the distance to the mesh or the closest sphere.
In practice, this means:

\begin{equation}
\phi_{\mathrm{mesh}}(x) =
\begin{cases}
d_{\mathrm{mesh}}(x), & \text{if } x \text{ is inside the mesh},\\[0.4em]
0, & \text{if } x \text{ is on the surface},\\[0.4em]
-\infty, & \text{if } x \text{ is outside the mesh}.
\end{cases}
\end{equation}

Here, \(d_{\mathrm{mesh}}(x)\) denotes the shortest Euclidean distance from \(x\) to the mesh surface. The program works with a rough grid and a refined grid based on the best rough grid point. 

\begin{figure}[t]
    \centering
    \includegraphics[width=8.9cm]{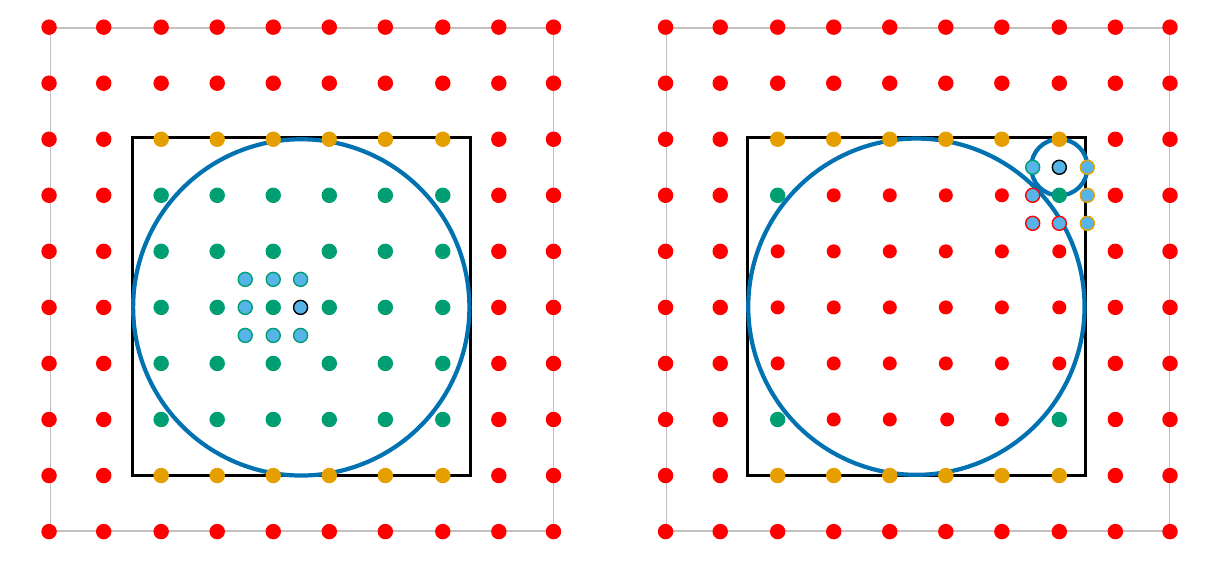}
    \caption{This figure displays two situations, the placement of the first sphere (left), and the placement of the second sphere (right). It can be seen that in the workspace \(\Omega\) (Grey), a grid of feasible points (Green), and infeasible points (Red and Orange), change after the first sphere is placed.}
    \label{fig:Appendix:MDBD}
\end{figure}

To give an example, consider a cube-shaped mesh positioned in the workspace. Inside the cube, the interior distance values are positive; they gradually decrease to zero on the cube’s surface and are masked to negative infinity outside of it. In Fig.~\ref{fig:Appendix:MDBD} two consecutive situations are displayed. On the left, the placement of the first sphere is seen and on the right the placement of the second sphere. The border of workspace \(\Omega\) is Grey, the \(-\infty\) nodes are red, the nodes that are zero are orange, the positive nodes are green, and the refined grid points are light blue with a red, orange, green, or black outline based on their feasibility, with the black outline being the optimum. The watertight mesh \(\Phi_{\mathrm{mesh}}\) is black and the spheres are blue.

Now it can be seen that on the left panel of Fig.~\ref{fig:Appendix:MDBD} that a rough grid node is surrounded by refined grid points. From this refined grid the largest interior distance is chosen and the sphere is placed in the center of the cube. Then all the nodes inside of the sphere are marked as infeasible and a new optimum point is found to place the next sphere, which can be seen in the right panel of Fig.~\ref{fig:Appendix:MDBD}.

To achieve this, first the distance from each node to the closest mesh point \(d_{\mathrm{mesh}}(x_{ijk})\) is determined by using the OPEN3D toolbox in Python. This provides a complete list of all points with the corresponding distances to the closest mesh point. After which the points are "masked": the points outside the mesh become \(-\infty\), the points on the border become zero, and the distances on the inside become positive. 

The most interior point, which corresponds to the center location where the largest sphere can be placed, can then be found using:
\begin{equation}
(i^\star,j^\star,k^\star) = \arg\max_{i,j,k}\phi_{\mathrm{mesh}}(x_{ijk}), 
\qquad 
r^\star = \max_{i,j,k}\phi_{\mathrm{mesh}}(x_{ijk}).
\end{equation}

This works for the first sphere. However, when at least one sphere is present, this no longer works. When a sphere is placed, an additional distance is connected to each node, \(d_{\mathrm{sph}}(x_{ijk})\), which is determined as the distance to the closest existing sphere:
\begin{equation}
d_{\mathrm{sph}}(x_{ijk}) =
\begin{cases}
+\infty, & \text{if no spheres are placed},\\[0.4em]
\min\limits_m \big(\|x_{ijk} - c_m\|_2 - r_m\big), & \text{otherwise}.
\end{cases}
\end{equation}
Here, \(c_m\) is the center of a sphere and \(r_m\) its radius. The distance \(d_{\mathrm{sph}}(x_{ijk})\) indicates the distance of a node to the closest sphere. If the distance becomes negative, as a node lies inside of a sphere, the node is masked with minus infinity, as can be seen on the right in Fig.~\ref{fig:Appendix:MDBD}.

Now, each node has two distance values. We are interested in the smallest value as that indicates the distance of the closest object to the node, either mesh or sphere:
\begin{equation}
\phi_{\mathrm{comb}}(x_{ijk}) = 
\min\big(\phi_{\mathrm{mesh}}(x_{ijk}),\, d_{\mathrm{sph}}(x_{ijk})\big).
\end{equation}

The maximum of this combined field gives the next sphere center and its radius:
\begin{equation}
(i^\star,j^\star,k^\star) = \arg\max_{i,j,k}\phi_{\mathrm{comb}}(x_{ijk}), 
\qquad 
r^\star = \max_{i,j,k}\phi_{\mathrm{comb}}(x_{ijk}).
\end{equation}

The same method can be applied to both the rough grid and the refined grid. When the optimal point is found, a sphere can be placed with radius \(r_m = \phi_{\mathrm{comb}}(x_{i^\star j^\star k^\star})\), as this distance represents the maximum radius possible without intersection. By looping in this manner for a desired number of spheres, the algorithm produces the Maximal Disjoint Ball Decomposition using a brute-force method. The pseudo-code of the algorithm can be seen in Algorithm~\ref{alg:mdbd_bruteforce}.

\begin{algorithm}
\caption{MDBD\_BruteForce($\Phi_{\mathrm{mesh}}$, $\Omega$, $N_r$, $N_f$, $K$)}
\label{alg:mdbd_bruteforce}
\KwIn{Watertight mesh $\Phi_{\mathrm{mesh}}$, workspace bounds $\Omega$, coarse grid resolution $N_r$, fine patch size $N_f$, max spheres $K$}
\KwOut{List of disjoint spheres}

\BlankLine
\textbf{1. Coarse grid setup and mesh distance field}\\
Generate a rectilinear coarse grid $G_r=\{x_{ijk}\}\subset\Omega$ with $N_r$ nodes per axis.\\
Compute $d_{\mathrm{mesh}}(x)$ using OPEN3D (closest point to $\Phi_{\mathrm{mesh}}$).\\
Mask: set $\phi_{\mathrm{mesh}}(x)=-\infty$ for outside, $\phi_{\mathrm{mesh}}(x)=0$ on surface, and $\phi_{\mathrm{mesh}}(x)>0$ inside.\\
Initialize $d_{\mathrm{sph}}(x)\leftarrow +\infty$ on $G_r$.

\BlankLine
\textbf{2. Iterative placement (coarse $\rightarrow$ fine)}\\
\For{$m=1$ \KwTo $K$}{
  Compute $\phi_{\mathrm{comb}}(x)=\min(\phi_{\mathrm{mesh}}(x),d_{\mathrm{sph}}(x))$.\\
  Find $x^\star=\arg\max\phi_{\mathrm{comb}}(x)$ and $r^\star_{\mathrm{coarse}}=\max\phi_{\mathrm{comb}}(x)$.\\
  \If{$r^\star_{\mathrm{coarse}}\le0$}{\textbf{break}}

  Construct local fine grid $G_f$ around $x^\star$.\\
  Recompute $\phi_{\mathrm{mesh}}(p)$ and $d_{\mathrm{sph}}(p)$ on $G_f$.\\
  Compute $\phi_{\mathrm{comb}}(p)=\min(\phi_{\mathrm{mesh}}(p),d_{\mathrm{sph}}(p))$.\\
  Find $p^\star=\arg\max\phi_{\mathrm{comb}}(p)$ and $r^\star=\max\phi_{\mathrm{comb}}(p)$.\\
  \If{$r^\star\le0$}{\textbf{continue}}

  Append $(c_m,r_m)\leftarrow(p^\star,r^\star)$ to sphere list.\\
  Update $d_{\mathrm{sph}}(x)\leftarrow \min\bigl(d_{\mathrm{sph}}(x),\,\|x-c_m\|_2-r_m\bigr)$ for all $x\in G_r$.\\
  \textit{(Do not modify $\phi_{\mathrm{mesh}}$; sphere interiors are already excluded via $d_{\mathrm{sph}}<0$ in $\phi_{\mathrm{comb}}$.)}
}

\BlankLine
\Return MDBD sphere list
\end{algorithm}

\subsection{Test cases for the MDBD Algorithm}
To test the functionality of the Brute Force MDBD algorithm two tests are conducted. The first one is placing 14 spheres into a $2\times1\times1$ cuboid and checking if the spheres are placed as expected, do not exceed the boundary box, or overlap with each other. The expected result is two large spheres with a radius of approximately 0.5, eight spheres on the corners of the cuboid and four spheres around the center.

The second test compares the speed and accuracy of this brute-force algorithm with the marching cubes (greedy) approach presented in \cite{Vu}. For this comparison, a $1270\times1270\times2540\,\mathrm{mm}$ cuboid is filled with one sphere. The placed position is compared in accuracy to the theoretical optimal position at $(635, 635, 1905)$. The mean percentage error is compared to the computation time for placing the sphere. The mean percentage error is calculated as:

\begin{equation}
\mathrm{err}_{\%} = 
\frac{100}{n_\mathrm{d}}\sum_{k=1}^{n_\mathrm{d}}
\frac{|c^{(k)}_{\mathrm{comp}} - c^{(k)}_{\mathrm{theo}}|}{|c^{(k)}_{\mathrm{theo}}|},
\qquad n_\mathrm{d}=3.
\end{equation}

This follows the same calculation used in \cite{Vu}. The resolutions of the coarse and fine grids are varied and the results compared accordingly.

\subsection{Results}
With evaluating the results generated by the Brute-Force algorithm we can see if this method can be used to generate MDBD objects for the optimization algorithm presented in this paper. In addition, by comparing the Brute-Force algorithm to the marching cube algorithm, we provide a comparison between the two MDBD algorithms.
\subsubsection{2x1x1 cuboid with 14 Spheres}
First, the results of the cuboid filled with 14 spheres.
\begin{table}[h!]
\centering
\caption{Sphere centers and radii within the $2\times1\times1$ cuboid.}
\label{tab:SphereResult}
\begin{tabularx}{\columnwidth}{@{}c *{4}{>{\centering\arraybackslash}X}@{}}
\toprule
\textbf{Sphere} & \textbf{Center X} & \textbf{Center Y} & \textbf{Center Z} & \textbf{Radius} \\
\midrule
1  & 0.500031 & 0.500000 & 0.500000 & 0.500000 \\
2  & 1.499969 & 0.499938 & 0.499938 & 0.499938 \\
3  & 1.000000 & 0.177103 & 0.177103 & 0.177103 \\
4  & 1.000000 & 0.177103 & 0.822897 & 0.177103 \\
5  & 1.000000 & 0.822897 & 0.177103 & 0.177103 \\
6  & 1.000000 & 0.822897 & 0.822897 & 0.177103 \\
7  & 1.865894 & 0.865987 & 0.865987 & 0.134005 \\
8  & 1.866018 & 0.134013 & 0.865987 & 0.133982 \\
9  & 1.866018 & 0.865987 & 0.134013 & 0.133982 \\
10 & 0.133982 & 0.133951 & 0.133951 & 0.133951 \\
11 & 0.133982 & 0.133951 & 0.865987 & 0.133951 \\
12 & 0.133982 & 0.865987 & 0.133951 & 0.133951 \\
13 & 0.133982 & 0.865987 & 0.866049 & 0.133951 \\
14 & 1.866018 & 0.133951 & 0.133951 & 0.133951 \\
\bottomrule
\end{tabularx}
\end{table}

\begin{figure}[t]
    \centering
    \includegraphics[width=6cm,angle=-90]{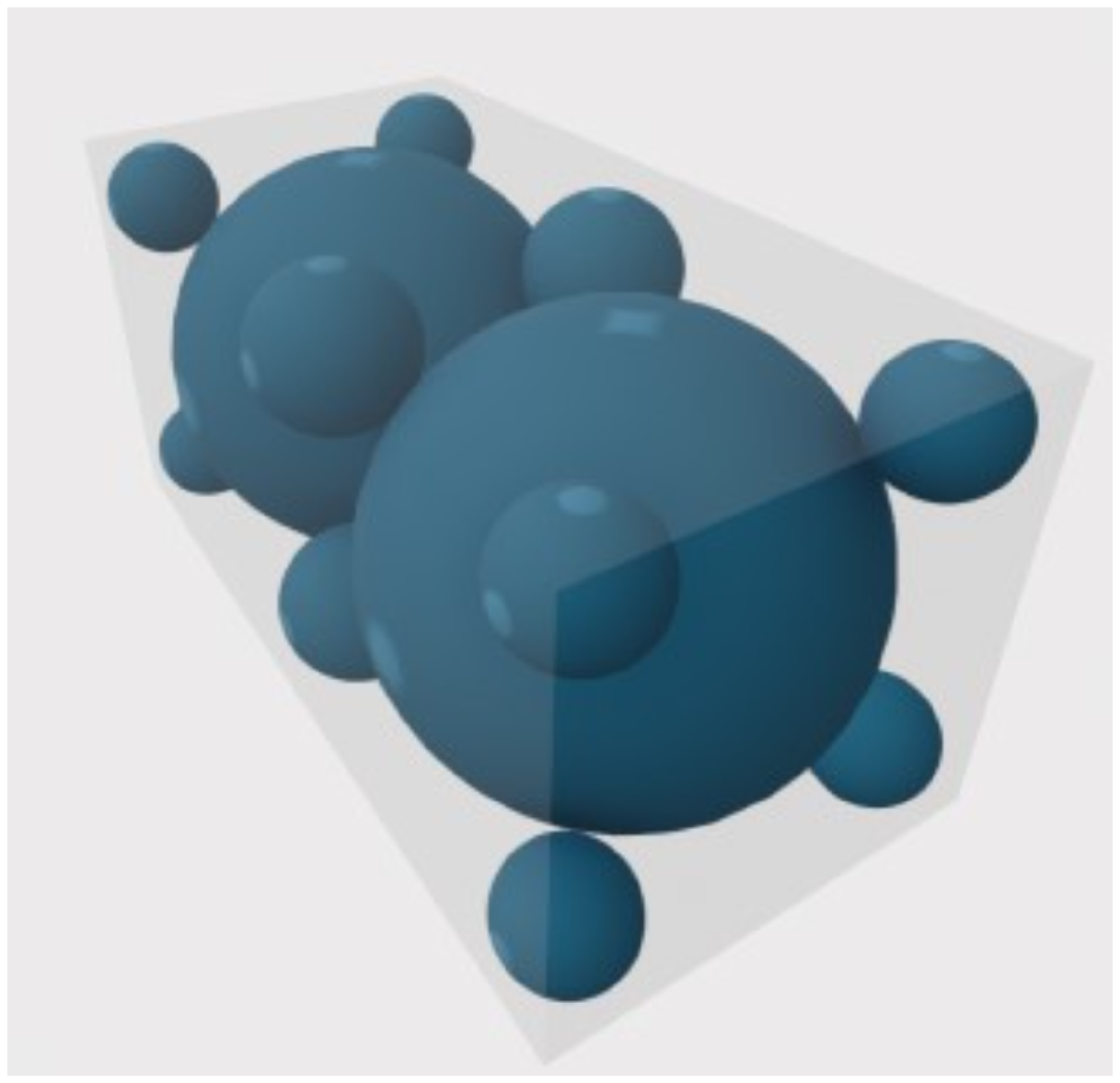}
    \caption{14 Spheres placed in a 2x1x1 Cuboid}
    \label{fig:Appendix:A3}
\end{figure}

From table \ref{tab:SphereResult} it can be seen that the spheres are placed as expected, in addition it can also be seen that no spheres overlap or exceed the boundary box. This can also be seen in Fig.~\ref{fig:Appendix:A3}, however it must be noted that the more light blue areas on the tangency of the spheres to the box is merely a visual effect from the 3D tool used to visualize the results. The spheres do not interfere with the boundary box or each other. 

\subsubsection{Comparison}
\begin{figure}[t]
    \centering
    \includegraphics[width=8.9cm]{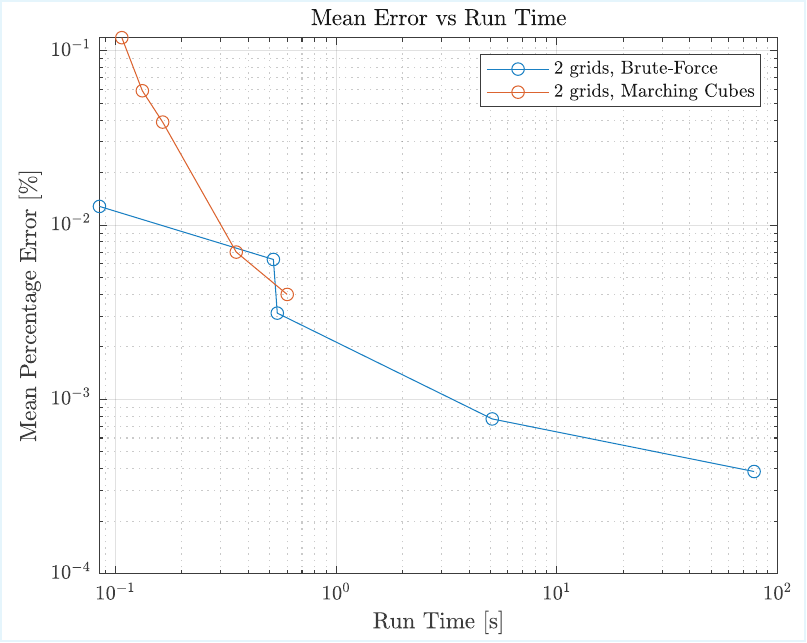}
    \caption{Comparison of Brute-Force and Marching Cubes
performance and accuracy.}
    \label{fig:Appendix:A1}
\end{figure}

When comparing the marching cube approach from \cite{Vu} to the brute-force method developed for this paper,  the numerical results can be seen in Table~\ref{tab:BFvsMC} and this table is also displayed in Fig.~\ref{fig:Appendix:A1}. From the results we can see that the Brute-Force algorithm is more accurate, but also slower than the Marching Cube Algorithm. 

\begin{table}[h!]
\centering
\caption{Comparison of Brute-Force and Marching Cubes performance and accuracy.}
\label{tab:BFvsMC}
\begin{tabularx}{\columnwidth}{@{}l *{4}{>{\centering\arraybackslash}X}@{}}
\toprule
\textbf{Grid Size (Rough/Fine)} & \textbf{Brute-Force Time (s)} & \textbf{Brute-Force Error} & \textbf{Marching Cubes Time (s)} & \textbf{Marching Cubes Error} \\
\midrule
32/64   & 0.0846 & 0.0128\% & 0.1069 & 0.119\% \\
64/128  & 0.5410 & 0.0031\% & 0.1636 & 0.039\% \\
32/128  & 0.5194 & 0.0064\% & 0.1323 & 0.059\% \\
128/256 & 5.0941 & 0.0007\% & 0.3526 & 0.007\% \\
128/512 & 78.301 & 0.0004\% & 0.5999 & 0.004\% \\
\bottomrule
\end{tabularx}
\end{table}

\subsection{Conclusion}
To restate the intention of this section, the aim is to develop an algorithm to generate MDBD objects from mesh files. When looking at the results, we can conclude that the program functions as desired. Based on the comparison to \cite{Vu}, we can also conclude that the program works slower but more accurate for equal rough grid and fine grid resolution. Going forward in this paper a rough grid of 64 is used with a fine grid of 128, as this provides a good tradeoff between speed and accuracy.

\newpage
\section{}\label{Appendix:B}
In this appendix a more elaborate explanation of the routing–object constraint and the routing–routing constraint can be found. First the Routing–Object interference calculation will be elaborated on, then the Routing–Routing interference will be explained.

\subsection{Rotation Matrices}
We use an intrinsic roll–pitch–yaw (XYZ) composition:
\begin{equation}
\mathbf{R}_i=\mathbf{R}_x(x_{\theta,i})\,\mathbf{R}_y(x_{\alpha,i})\,\mathbf{R}_z(x_{\beta,i}),
\end{equation}
with the direction cosine matrices;
\begin{align}
\mathbf{R}_\mathrm{x}(\theta) &=
\begin{bmatrix}
1 & 0 & 0\\
0 & \cos\theta & -\sin\theta\\
0 & \sin\theta & \cos\theta
\end{bmatrix} \\[6pt]
\mathbf{R}_\mathrm{y}(\alpha) &=
\begin{bmatrix}
\cos\alpha & 0 & \sin\alpha\\
0 & 1 & 0\\
-\sin\alpha & 0 & \cos\alpha
\end{bmatrix} \\[6pt]
\mathbf{R}_\mathrm{z}(\beta) &=
\begin{bmatrix}
\cos\beta & -\sin\beta & 0\\
\sin\beta & \cos\beta & 0\\
0 & 0 & 1
\end{bmatrix}.
\end{align}

\subsection{Routing–Object Interference}
First, we restate and then continue the mathematical explanation from the Methods section. 
Each route $L$ consists of $K_L$ straight segments that connect $K_L+1$ nodes 
$[\mathbf{q}_{L,0},\,\mathbf{q}_{L,1},\,\dots,\,\mathbf{q}_{L,K_L}]$, 
where the endpoints $\mathbf{q}_{L,0}$ and $\mathbf{q}_{L,K_L}$ are ports and the intermediate nodes 
$\mathbf{q}_{L,k}$ ($k=1,\dots,K_L-1$) are control points $\mathbf{c}_{L,k}$. 
The number of segments is $K_L = n_{\mathrm{cp},L}+1$, where $n_{\mathrm{cp},L}$ is the number of control points in route $L$.
Each segment $m$ connects nodes $\mathbf{q}_{L,m}$ and $\mathbf{q}_{L,m+1}$ for $m=0,\dots,K_L-1$. 
The routing tube has constant radius $r_r$.
Each object $A_i$ consists of spheres $b_{i,\mu}$ with centers 
$\mathbf{p}^{\mathbb{W}}_{\mathrm{b}_{i,\mu}}\in\mathbb{R}^3$ and radii $r_{\mathrm{b}_{i,\mu}}$.

For a segment $[\mathbf{q}_{L,m},\,\mathbf{q}_{L,m+1}]$ and a sphere center 
$\mathbf{p}^{\mathbb{W}}_{\mathrm{b}_{i,\mu}}$, define
\begin{align*}
\mathbf{a}_{L,m} &= \mathbf{q}_{L,m},\\
\mathbf{b}_{L,m} &= \mathbf{q}_{L,m+1},\\
\mathbf{u}_{L,m} &= \mathbf{b}_{L,m}-\mathbf{a}_{L,m}.
\end{align*}
The projection parameter of the sphere center onto the line defined by the segment is
\begin{align*}
t^*_{L,m,i,\mu} &= 
\frac{(\mathbf{p}^{\mathbb{W}}_{\mathrm{b}_{i,\mu}}-\mathbf{a}_{L,m})^\top \mathbf{u}_{L,m}}
{\mathbf{u}^\top_{L,m}\mathbf{u}_{L,m}+\varepsilon},
\quad \varepsilon>0,
\end{align*}
and the clamped parameter on the segment is
\begin{align*}
t_{L,m,i,\mu} &= \min\bigl(\max(t^*_{L,m,i,\mu},0),\,1\bigr).
\end{align*}
The corresponding closest point on the segment is then
\begin{align*}
\mathbf{p}_{L,m,i,\mu} &= \mathbf{a}_{L,m} + t_{L,m,i,\mu}\,\mathbf{u}_{L,m}.
\end{align*}
The minimum clearance between routing segment $(L,m)$ and sphere $b_{i,\mu}$ is
\[
d^{\,\mathrm{route\mbox{-}obj}}_{L,m,i,\mu}
= \bigl\|\mathbf{p}^{\mathbb{W}}_{\mathrm{b}_{i,\mu}}-\mathbf{p}_{L,m,i,\mu}\bigr\|_2
  - \bigl(r_{\mathrm{b}_{i,\mu}} + r_r\bigr).
\]
To avoid interference, $d^{\,\mathrm{route\mbox{-}obj}}_{L,m,i,\mu} \ge 0$ must hold for all routes $L$, segments $m$, objects $i$, and spheres $\mu$.

Stacking all routing–object clearances yields
\[
\mathbf{g}^{\mathrm{route\mbox{-}obj}}
= -\bigl[d^{\,\mathrm{route\mbox{-}obj}}_{L,m,i,\mu}\bigr]_{(L,m,i,\mu)}
\in \mathbb{R}^{N_{\mathrm{r\mbox{-}o}}},
\]
where the total number of constraint equations is
\[
N_{\mathrm{r\mbox{-}o}} =
\sum_{L}\sum_{m=0}^{K_L-1}\sum_{i=1}^{n_\mathrm{obj}} n_{\mathrm{b},i}.
\]
Finally, the constraints are enforced in negative-null form as
\begin{equation}
\mathbf{g}^{\mathrm{route\mbox{-}obj}}(\mathbf{x}) \le 0.
\end{equation}

\subsection*{Routing–Routing Interference}
\label{app:routing-routing}
\noindent
Let segment $(L,m)$ have endpoints 
$\mathbf{a}_{L,m}=\mathbf{q}_{L,m}$, 
$\mathbf{b}_{L,m}=\mathbf{q}_{L,m+1}$, 
and direction vector 
$\mathbf{u}_{L,m}=\mathbf{b}_{L,m}-\mathbf{a}_{L,m}$.
Similarly, segment $(L',\eta)$ has endpoints 
$\mathbf{a}_{L',\eta}=\mathbf{q}_{L',\eta}$, 
$\mathbf{b}_{L',\eta}=\mathbf{q}_{L',\eta+1}$, 
and direction 
$\mathbf{v}_{L',\eta}=\mathbf{b}_{L',\eta}-\mathbf{a}_{L',\eta}$.

Define the auxiliary quantities:
\begin{align*}
\mathbf{w}_0 &= \mathbf{a}_{L,m} - \mathbf{a}_{L',\eta}\\, 
a &= \mathbf{u}_{L,m}^\top \mathbf{u}_{L,m} + \varepsilon, && \varepsilon>0,\\
b &= \mathbf{u}_{L,m}^\top \mathbf{v}_{L',\eta},\\
c &= \mathbf{v}_{L',\eta}^\top \mathbf{v}_{L',\eta} + \varepsilon,\\
d &= \mathbf{u}_{L,m}^\top \mathbf{w}_0,\\
e &= \mathbf{v}_{L',\eta}^\top \mathbf{w}_0,\\
D &= a\,c - b^2.
\end{align*}

The small constant $\varepsilon>0$ ensures numerical robustness in degenerate or nearly parallel cases.

The unconstrained closest-point parameters on the infinite lines are
\[
s^*=\frac{b\,e - c\,d}{D},\qquad
t^*=\frac{a\,e - b\,d}{D}.
\]
Clamping to the segment domains yields
\[
s=\min\bigl(\max(s^*,0),1\bigr),\qquad
t=\min\bigl(\max(t^*,0),1\bigr).
\]
The corresponding closest points on the two segments are
\[
\mathbf{p}_{L,m}=\mathbf{a}_{L,m}+s\,\mathbf{u}_{L,m},\qquad
\mathbf{p}_{L',n}=\mathbf{a}_{L',\eta}+t\,\mathbf{v}_{L',\eta}.
\]
The minimum clearance (tube gap) between the two routing segments is
\[
d^{\,\mathrm{route\mbox{-}route}}_{L,m,L',\eta}
= \bigl\|\mathbf{p}_{L,m}-\mathbf{p}_{L',\eta}\bigr\|_2
- 2\,r_r.
\]
Non-interference requires
\[
d^{\,\mathrm{route\mbox{-}route}}_{L,m,L',\eta}\ge 0
\quad\forall\ (L,m,L',\eta)\in\mathcal{C},
\]
where the set of evaluated segment pairs is defined as
\begin{equation}
\begin{aligned}
\mathcal{C}
=\Bigl\{(L,m,L',\eta)\ \big|\
\big(L<L',\ 0\le m<K_L,\ 0\le \eta<K_{L'}\big)\\
\text{or}\ 
\big(L=L',\ 0\le m<\eta<K_L,\ |m-\eta|\ge 2\big)
\Bigr\}.
\end{aligned}
\label{eq:SetC}
\end{equation}

The exclusion of pairs with $L=L'$ and $|m-\eta|\le1$ allows consecutive segments of the same route to meet at their shared endpoint without triggering a false self-collision. 
\newpage
\section{}\label{Appendix:C}
This appendix provides a detailed explanation of the iterative update rules used in the Analytical Target Cascading (ATC) framework described in Section~\ref{section:Methods}. 
While the main body of the paper introduced the hierarchical formulation and optimization problems at both the system and subsystem levels, this appendix elaborates on the parameter update mechanisms that enable convergence between levels. 
Specifically, it details how the target variables, dual variables, coupling parameters, and penalty weights are updated during each ATC iteration to ensure consistency and feasibility across the coupled optimization hierarchy. The equations used are similar to \cite{ATC, ATC2, ATC3}.

After each iteration, two updates are performed.  
First, the new targets for each subsystem are updated according to a relaxation rule:
\[
\boldsymbol{\gamma}_i^{(k+1)} 
= (1 - s)\,\boldsymbol{\gamma}_i^{(k)} 
  + s\,\mathbf{x}_{\mathrm{sys},i}^{(k)},
\]
where \(s \in (0,1]\) is the target step size controlling the update rate 
(\(s=1\) gives a full update).  

Second, the dual variables are updated as
\[
\boldsymbol{\lambda}_i^{(k+1)} 
= \boldsymbol{\lambda}_i^{(k)} 
  + \rho \bigl(\mathbf{x}^L_i - \boldsymbol{\gamma}_i^{(k)}\bigr),
\]
and the coupling parameter \(\pi\) is increased if constraint violations remain.  
The procedure continues until both the consistency norm 
\(\|\mathbf{x}^L_i - \boldsymbol{\gamma}_i\|\) 
and the constraint violations fall below the specified tolerances.

In this implementation, the coupling parameter \(\pi\) 
(denoted \(\eta\) in the system-level solver) 
is dynamically adapted to strengthen consistency between 
the system and subsystem solutions. 
After each iteration, if the system-level constraint function \(g(\mathbf{x})\) 
still violates feasibility (\(g(\mathbf{x}) < 0\)), 
the coupling parameter is increased according to
\[
\pi^{(k+1)} = \pi^{(k)} \times \pi_{\mathrm{growth}},
\]
where \(\pi_{\mathrm{growth}}\) typically ranges between 1.5 and 3.0.  
Increasing \(\pi\) enforces tighter coordination between levels, 
causing the system-level solution to align more closely 
with the subsystem responses in subsequent iterations.

The penalty weight \(\rho\), on the other hand, acts locally within each subsystem 
and determines how strongly it adheres to its assigned target. 
In the current implementation, \(\rho\) is fixed per subsystem. 
However, in extended formulations of ATC, it can be updated similarly according to
\[
\rho^{(k+1)} =
\begin{cases}
\rho^{(k)} \times \rho_{\mathrm{growth}}, & 
\text{if } \|\mathbf{x}^L_i - \boldsymbol{\gamma}_i^{(k)}\| > \text{tol},\\[3pt]
\rho^{(k)}, & \text{otherwise},
\end{cases}
\]
which tightens local consistency enforcement when subsystem deviations remain large.

\newpage
\onecolumn
\section{}\label{Appendix:E}
The use-case introduced by \cite{Behzadi} consists of a \(1.5 \times 1.5 \times 1.5\) MDBD Cube.  
The MDBD cube used in this paper is generated using the algorithm described in Appendix~\ref{Appendix:A}.  
The full MDBD geometry is provided in Table~\ref{tab:100Spheres}, while the corresponding port position, located at the center of the cube, is listed in Table~\ref{tab:100SpheresPort}.

\vspace{0.5cm}
\noindent
The routing connectivity for various configurations is defined below.

\paragraph{Three objects}
\[
\mathrm{connections} =
\big[
(\varphi_{1,1}, \varphi_{2,1}),
(\varphi_{2,1}, \varphi_{3,1}),
(\varphi_{3,1}, \varphi_{1,1})
\big]
\]

\paragraph{Four objects}
\[
\mathrm{connections} =
\big[
(\varphi_{1,1}, \varphi_{2,1}),
(\varphi_{1,1}, \varphi_{3,1}),
(\varphi_{2,1}, \varphi_{4,1}),
(\varphi_{3,1}, \varphi_{4,1})
\big]
\]

\paragraph{Six objects}
\[
\mathrm{connections} =
\big[
(\varphi_{2,1}, \varphi_{3,1}),
(\varphi_{3,1}, \varphi_{1,1}),
(\varphi_{4,1}, \varphi_{2,1}),
(\varphi_{1,1}, \varphi_{4,1}),
(\varphi_{6,1}, \varphi_{1,1}),
(\varphi_{4,1}, \varphi_{5,1}),
(\varphi_{3,1}, \varphi_{5,1}),
(\varphi_{6,1}, \varphi_{5,1})
\big]
\]

\begin{longtable}[c]{rrrr}
\caption{Sphere centers and radii (x, y, z, radius)}\label{tab:100Spheres} \\
\toprule
$x$ & $y$ & $z$ & Radius \\
\midrule
\endfirsthead

\toprule
$x$ & $y$ & $z$ & Radius \\
\midrule
\endhead

-6.94E-17 & -6.94E-17 & -6.94E-17 & 0.75 \\
-0.54911886 & -0.54911886 & -0.54911886 & 0.20088114 \\
-0.55568054 & -0.55568054 & 0.523622047 & 0.19431946 \\
-0.55568054 & 0.523622047 & -0.55568054 & 0.19431946 \\
-0.55568054 & 0.523622047 & 0.55568054 & 0.19431946 \\
0.523622047 & -0.55568054 & -0.55568054 & 0.19431946 \\
0.523622047 & -0.55568054 & 0.55568054 & 0.19431946 \\
0.523622047 & 0.55568054 & -0.55568054 & 0.19431946 \\
0.523622047 & 0.55568054 & 0.55568054 & 0.19431946 \\
-0.608361455 & -0.23472066 & 0.608173978 & 0.141638545 \\
-0.608361455 & 0.608173978 & -0.23472066 & 0.141638545 \\
-0.608361455 & 0.608173978 & 0.23472066 & 0.141638545 \\
-0.23472066 & -0.608361455 & 0.608173978 & 0.141638545 \\
-0.23472066 & 0.608173978 & -0.608361455 & 0.141638545 \\
-0.23472066 & 0.608173978 & 0.608361455 & 0.141638545 \\
0.608173978 & -0.608361455 & -0.23472066 & 0.141638545 \\
0.608173978 & -0.608361455 & 0.23472066 & 0.141638545 \\
0.608173978 & -0.23472066 & -0.608361455 & 0.141638545 \\
0.608173978 & -0.23472066 & 0.608361455 & 0.141638545 \\
0.608173978 & 0.23472066 & -0.608361455 & 0.141638545 \\
0.608173978 & 0.23472066 & 0.608361455 & 0.141638545 \\
0.608173978 & 0.608361455 & -0.23472066 & 0.141638545 \\
0.608173978 & 0.608361455 & 0.23472066 & 0.141638545 \\
-0.610048744 & -0.610048744 & -0.219347582 & 0.139951256 \\
-0.610048744 & -0.219347582 & -0.610048744 & 0.139951256 \\
-0.219347582 & -0.610048744 & -0.610048744 & 0.139951256 \\
-0.611736033 & -0.611736033 & 0.200787402 & 0.138105938 \\
-0.611736033 & 0.200787402 & -0.611736033 & 0.138105938 \\
-0.611736033 & 0.200787402 & 0.611736033 & 0.138105938 \\
0.200787402 & -0.611736033 & -0.611736033 & 0.138105938 \\
0.200787402 & -0.611736033 & 0.611736033 & 0.138105938 \\
0.200787402 & 0.611736033 & -0.611736033 & 0.138105938 \\
0.200787402 & 0.611736033 & 0.611736033 & 0.138105938 \\
-0.649981252 & 0.547619048 & -0.004499438 & 0.099930876 \\
0.547619048 & -0.649981252 & -0.004499438 & 0.099930876 \\
0.547619048 & -0.004499438 & -0.649981252 & 0.099930876 \\
0.547619048 & -0.004499438 & 0.649981252 & 0.099930876 \\
0.547619048 & 0.649981252 & -0.004499438 & 0.099930876 \\
-0.533933258 & -0.014810649 & 0.654855643 & 0.09506804 \\
-0.014810649 & -0.533933258 & 0.654855643 & 0.09506804 \\
-0.014810649 & 0.654855643 & -0.533933258 & 0.09506804 \\
-0.014810649 & 0.533933258 & 0.654855643 & 0.09506804 \\
-0.404761905 & -0.347206599 & 0.655230596 & 0.094745634 \\
-0.404761905 & 0.655230596 & -0.347206599 & 0.094745634 \\
-0.404761905 & 0.655230596 & 0.347206599 & 0.094745634 \\
0.655230596 & -0.404761905 & -0.347206599 & 0.094745634 \\
0.655230596 & -0.404761905 & 0.347206599 & 0.094745634 \\
0.655230596 & 0.347206599 & -0.404761905 & 0.094745634 \\
0.655230596 & 0.347206599 & 0.404761905 & 0.094745634 \\
-0.657667792 & -0.526059243 & -0.008061492 & 0.092217453 \\
-0.657667792 & -0.008061492 & -0.526059243 & 0.092217453 \\
-0.526059243 & -0.657667792 & -0.008061492 & 0.092217453 \\
-0.526059243 & -0.008061492 & -0.657667792 & 0.092217453 \\
-0.008061492 & -0.657667792 & -0.526059243 & 0.092217453 \\
-0.008061492 & -0.526059243 & -0.657667792 & 0.092217453 \\
-0.660104987 & -0.409261342 & -0.319647544 & 0.089868117 \\
-0.409261342 & -0.660104987 & -0.319647544 & 0.089868117 \\
-0.409261342 & -0.319647544 & -0.660104987 & 0.089868117 \\
-0.659730034 & -0.026809149 & 0.519497563 & 0.090109179 \\
-0.026809149 & -0.659730034 & 0.519497563 & 0.090109179 \\
-0.026809149 & 0.519497563 & -0.659730034 & 0.090109179 \\
-0.026809149 & 0.659730034 & 0.519497563 & 0.090109179 \\
-0.415823022 & 0.3023997 & 0.661792276 & 0.088047362 \\
0.3023997 & -0.415823022 & 0.661792276 & 0.088047362 \\
0.3023997 & 0.415823022 & 0.661792276 & 0.088047362 \\
0.3023997 & 0.661792276 & -0.415823022 & 0.088047362 \\
0.3023997 & 0.661792276 & 0.415823022 & 0.088047362 \\
-0.661792276 & -0.415823022 & 0.3023997 & 0.088047362 \\
-0.661792276 & 0.3023997 & -0.415823022 & 0.088047362 \\
-0.661792276 & 0.3023997 & 0.415823022 & 0.088047362 \\
-0.415823022 & -0.661792276 & 0.3023997 & 0.088047362 \\
-0.415823022 & 0.3023997 & -0.661792276 & 0.088047362 \\
0.3023997 & -0.661792276 & -0.415823022 & 0.088047362 \\
0.3023997 & -0.661792276 & 0.415823022 & 0.088047362 \\
0.3023997 & -0.415823022 & -0.661792276 & 0.088047362 \\
0.3023997 & 0.415823022 & -0.661792276 & 0.088047362 \\
-0.508998875 & 0.663104612 & 0.03655793 & 0.086734165 \\
0.663104612 & -0.508998875 & 0.03655793 & 0.086734165 \\
0.663104612 & 0.03655793 & -0.508998875 & 0.086734165 \\
0.663104612 & 0.03655793 & 0.508998875 & 0.086734165 \\
0.663104612 & 0.508998875 & 0.03655793 & 0.086734165 \\
-0.672290964 & -0.285901762 & 0.404761905 & 0.077639723 \\
-0.671728534 & -0.265091864 & -0.405699288 & 0.078271466 \\
-0.670416198 & 0.400449944 & -0.280089989 & 0.079583802 \\
-0.673415823 & 0.401574803 & 0.261717285 & 0.076584177 \\
-0.285901762 & -0.672290964 & 0.404761905 & 0.077639723 \\
-0.285901762 & 0.404761905 & -0.672290964 & 0.077639723 \\
-0.285901762 & 0.404761905 & 0.672290964 & 0.077639723 \\
-0.265091864 & -0.671728534 & -0.405699288 & 0.078271466 \\
-0.265091864 & -0.405699288 & -0.671728534 & 0.078271466 \\
0.400449944 & -0.670416198 & -0.280089989 & 0.079583802 \\
0.401574803 & -0.673415823 & 0.261717285 & 0.076584177 \\
0.400449944 & -0.280089989 & -0.670416198 & 0.079583802 \\
0.400449944 & -0.280089989 & 0.670416198 & 0.079583802 \\
0.401574803 & 0.261717285 & -0.673415823 & 0.076584177 \\
0.401574803 & 0.261717285 & 0.673415823 & 0.076584177 \\
0.400449944 & 0.670416198 & -0.280089989 & 0.079583802 \\
0.401574803 & 0.673415823 & 0.261717285 & 0.076584177 \\
-0.246531684 & -0.403449569 & 0.675478065 & 0.074366597 \\
-0.246531684 & 0.675478065 & -0.403449569 & 0.074366597 \\
\bottomrule
\end{longtable}

\begin{table}[h!]
\centering
\caption{Port coordinates (x, y, z, port nr)}\label{tab:100SpheresPort}
\begin{tabular}{rrrr}
\toprule
$x$ & $y$ & $z$ & Port nr \\
\midrule
0 & 0 & 0 & 1 \\
\bottomrule
\end{tabular}
\end{table}

\twocolumn
\clearpage
\newpage
\section{}\label{Appendix:F}
The MDBD compositions for the different object types are listed as follows: the Cuboid in Table~\ref{tab:Cuboid40SPheres} with its corresponding ports in Table~\ref{tab:portsCuboid}; the L-Shape in Table~\ref{tab:LShape40SPheres} with ports in Table~\ref{tab:portsLShape}; and the Double L-Shape (unique configuration) in Table~\ref{tab:DoubleLShape40SPheres} with ports in Table~\ref{tab:portsDoubleLShape}.
The objects are represented with 40 spheres. Configurations with 30 or 20 spheres can be derived by sequentially removing the smallest spheres until the desired number is reached.

\begin{table}[h!]
\centering
\small
\caption{Cuboid MDBD coordinates and radius for 40 spheres}
\label{tab:Cuboid40SPheres}
\begin{tabularx}{\columnwidth}{r|r|r|r}
\textbf{x} & \textbf{y} & \textbf{z} & \textbf{radius} \\
\hline
-0.4999 & -2.78E-17 & -2.78E-17 & 0.5000 \\
0.4999 & -2.78E-17 & -2.78E-17 & 0.4999 \\
-5.55E-17 & -0.32296 & -0.32296 & 0.1770 \\
-5.55E-17 & -0.32296 & 0.32283 & 0.1770 \\
-5.55E-17 & 0.32283 & -0.32296 & 0.1770 \\
-5.55E-17 & 0.32283 & 0.32296 & 0.1770 \\
-0.8659 & -0.36608 & -0.36608 & 0.1339 \\
-0.8704 & -0.37045 & 0.34908 & 0.1295 \\
-0.8704 & 0.34908 & -0.37045 & 0.1295 \\
-0.8704 & 0.34908 & 0.37045 & 0.1295 \\
0.8571 & -0.36833 & -0.36833 & 0.1317 \\
0.8571 & -0.36833 & 0.36833 & 0.1317 \\
0.8571 & 0.36833 & -0.36833 & 0.1317 \\
0.8571 & 0.36833 & 0.36833 & 0.1317 \\
-5.55E-17 & -0.37470 & -0.02499 & 0.1253 \\
-5.55E-17 & -0.02499 & -0.37470 & 0.1253 \\
-5.55E-17 & -0.02512 & 0.37470 & 0.1253 \\
-5.55E-17 & 0.37470 & -0.02512 & 0.1253 \\
0.26397 & -0.39433 & -0.39433 & 0.1056 \\
0.26397 & -0.39433 & 0.39433 & 0.1056 \\
0.26397 & 0.39433 & -0.39433 & 0.1056 \\
0.26397 & 0.39433 & 0.39433 & 0.1056 \\
-0.26422 & -0.39445 & -0.39445 & 0.1055 \\
-0.26422 & -0.39445 & 0.39445 & 0.1055 \\
-0.26422 & 0.39445 & -0.39445 & 0.1055 \\
-0.26422 & 0.39445 & 0.39445 & 0.1055 \\
0.9059 & -0.40607 & -0.15123 & 0.0939 \\
0.9059 & -0.40607 & 0.15123 & 0.0939 \\
0.9059 & -0.15123 & -0.40607 & 0.0939 \\
0.9059 & 0.15123 & -0.40607 & 0.0939 \\
0.9059 & -0.15123 & 0.40607 & 0.0939 \\
0.9059 & 0.15123 & 0.40607 & 0.0939 \\
0.9059 & 0.40607 & -0.15123 & 0.0939 \\
0.9059 & 0.40607 & 0.15123 & 0.0939 \\
-0.9054 & -0.15648 & 0.40557 & 0.0944 \\
-0.9054 & 0.40557 & -0.15648 & 0.0944 \\
-0.9054 & 0.40557 & 0.15648 & 0.0944 \\
-0.6564 & -0.40557 & 0.40545 & 0.0944 \\
-0.6564 & 0.40545 & -0.40557 & 0.0944 \\
-0.6564 & 0.40545 & 0.40557 & 0.0944 \\
\end{tabularx}
\end{table}

\begin{table}[h!]
\centering
\small
\caption{Port coordinates Cuboid}
\label{tab:portsCuboid}
\begin{tabularx}{0.6\columnwidth}{r|r|r|c}
\textbf{x} & \textbf{y} & \textbf{z} & \textbf{Port} \\
\hline
1.05 & 0 & 0 & 1 \\
-1.05 & 0 & 0 & 2 \\
\end{tabularx}
\end{table}

\begin{table}[h!]
\centering
\small
\caption{L-Shape MDBD coordinates and radius for 40 spheres}
\label{tab:LShape40SPheres}
\begin{tabularx}{\columnwidth}{r|r|r|r}
\textbf{x} & \textbf{y} & \textbf{z} & \textbf{radius} \\
\hline
-0.4999 & -0.4999 & -2.78E-17 & 0.5000 \\
-0.4999 & 0.4999 & -2.78E-17 & 0.4999 \\
0.4999 & -0.4999 & -2.78E-17 & 0.4999 \\
-0.1772 & -5.55E-17 & -0.32296 & 0.1770 \\
-0.1772 & -5.55E-17 & 0.32296 & 0.1770 \\
-0.8229 & -5.55E-17 & -0.32283 & 0.1771 \\
-0.8229 & -5.55E-17 & 0.32283 & 0.1771 \\
-5.55E-17 & -0.8229 & -0.32283 & 0.1771 \\
-5.55E-17 & -0.8229 & 0.32283 & 0.1771 \\
-5.55E-17 & -0.27422 & -0.34958 & 0.1504 \\
-5.55E-17 & -0.27422 & 0.34958 & 0.1504 \\
-0.8659 & -0.8659 & -0.36608 & 0.1339 \\
-0.8704 & -0.8704 & 0.34921 & 0.1296 \\
-0.8681 & 0.8571 & -0.36833 & 0.1317 \\
-0.8681 & 0.8571 & 0.36833 & 0.1317 \\
-0.1340 & 0.8659 & -0.36595 & 0.1340 \\
-0.1337 & 0.8661 & 0.36508 & 0.1337 \\
0.8571 & -0.8681 & -0.36833 & 0.1317 \\
0.8571 & -0.8681 & 0.36833 & 0.1317 \\
0.8659 & -0.1340 & -0.36595 & 0.1340 \\
0.8661 & -0.1337 & 0.36508 & 0.1337 \\
0.0045 & -0.1290 & -0.1111 & 0.1288 \\
-0.8746 & -5.55E-17 & -0.02487 & 0.1253 \\
-5.55E-17 & -0.8746 & -0.02487 & 0.1253 \\
-0.5249 & -5.55E-17 & -0.37470 & 0.1253 \\
-0.5249 & -5.55E-17 & 0.37470 & 0.1253 \\
0.0317 & -0.11799 & 0.13386 & 0.1178 \\
0.2097 & -0.11474 & -0.38095 & 0.1147 \\
0.2097 & -0.11474 & 0.38095 & 0.1147 \\
-0.1435 & 0.00025 & 0.03299 & 0.1148 \\
-0.02025 & -0.5374 & -0.38433 & 0.1156 \\
-0.02025 & -0.5374 & 0.38433 & 0.1156 \\
-0.1057 & 0.2640 & -0.39433 & 0.1056 \\
-0.1057 & 0.2640 & 0.39433 & 0.1056 \\
-0.9024 & -0.31746 & -0.40232 & 0.0976 \\
-0.9024 & -0.31746 & 0.40232 & 0.0976 \\
-0.8966 & 0.26022 & -0.39683 & 0.1032 \\
-0.8989 & 0.26122 & 0.38095 & 0.1011 \\
-0.31746 & -0.9024 & -0.40232 & 0.0976 \\
-0.31746 & -0.9024 & 0.40232 & 0.0976 \\
\end{tabularx}
\end{table}

\begin{table}[h!]
\centering
\small
\caption{Port coordinates L-Shape}
\label{tab:portsLShape}
\begin{tabularx}{0.6\columnwidth}{r|r|r|c}
\textbf{x} & \textbf{y} & \textbf{z} & \textbf{Port} \\
\hline
0.5 & -1.05 & 0 & 1 \\
-0.5 & 1.05 & 0 & 2 \\
\end{tabularx}
\end{table}

\begin{table}[h!]
\centering
\small
\caption{Double L-Shape MDBD coordinates and radius for 40 spheres}
\label{tab:DoubleLShape40SPheres}
\begin{tabularx}{\columnwidth}{r|r|r|r}
\textbf{x} & \textbf{y} & \textbf{z} & \textbf{r} \\
\hline
-0.4999 & -0.4999 & 6.25E-05 & 0.4999 \\
-0.4999 & -0.4999 & 0.9999 & 0.4999 \\
-0.4999 & 0.4999 & 6.25E-05 & 0.4999 \\
0.4999 & 0.4999 & 6.25E-05 & 0.4999 \\
-5.55E-17 & 0.1772 & 0.3228 & 0.1770 \\
-0.1772 & -0.1772 & 0.5 & 0.1770 \\
-0.8229 & -0.1772 & 0.5 & 0.1771 \\
-0.1772 & -0.8229 & 0.5 & 0.1771 \\
-0.1772 & -5.55E-17 & -0.3229 & 0.1771 \\
-5.55E-17 & 0.8229 & 0.3228 & 0.1771 \\
-0.8344 & -0.8344 & 0.4683 & 0.1656 \\
-0.8344 & -0.0317 & -0.3344 & 0.1656 \\
-0.0317 & 0.8344 & -0.3344 & 0.1656 \\
-5.55E-17 & 0.2742 & -0.3494 & 0.1505 \\
-0.8494 & -5.55E-17 & 0.2258 & 0.1505 \\
-0.8659 & -0.8659 & -0.3659 & 0.1339 \\
-0.8684 & -0.8684 & 1.3569 & 0.1316 \\
-0.8659 & -0.1340 & 1.3659 & 0.1339 \\
-0.8684 & 0.8569 & -0.3684 & 0.1316 \\
-0.8659 & 0.8659 & 0.3660 & 0.1339 \\
-0.1340 & -0.8659 & -0.3659 & 0.1339 \\
-0.1340 & -0.8659 & 1.3659 & 0.1339 \\
-0.1340 & -0.1340 & 1.3659 & 0.1339 \\
0.8569 & 0.8684 & -0.3684 & 0.1316 \\
0.8659 & 0.1340 & -0.3659 & 0.1339 \\
0.8659 & 0.1340 & 0.3660 & 0.1339 \\
0.8659 & 0.8659 & 0.3660 & 0.1339 \\
-0.3810 & 0.0092 & 0.3723 & 0.1273 \\
-0.1277 & -0.0092 & 0.1190 & 0.1273 \\
-0.8739 & -0.5469 & 0.5 & 0.1261 \\
-0.8749 & 0.0022 & -0.0487 & 0.1250 \\
-0.5469 & -0.8739 & 0.5 & 0.1261 \\
-0.5469 & -5.55E-17 & -0.3739 & 0.1261 \\
0.0022 & 0.5487 & -0.3749 & 0.1250 \\
-5.55E-17 & 0.8739 & -0.0469 & 0.1261 \\
-5.55E-17 & 0.1307 & -0.1074 & 0.1307 \\
-0.6254 & -5.55E-17 & 0.3725 & 0.1173 \\
-0.1252 & -0.5249 & 0.5 & 0.1252 \\
0.2110 & 0.1157 & -0.3844 & 0.1156 \\
-0.8844 & 0.2110 & 0.3843 & 0.1156 \\
\end{tabularx}
\end{table}

\begin{table}[h!]
\centering
\small
\caption{Port coordinates Double L-Shape}
\label{tab:portsDoubleLShape}
\begin{tabularx}{0.7\columnwidth}{r|r|r|c}
\textbf{x} & \textbf{y} & \textbf{z} & \textbf{Port} \\
\hline
-0.5 & -1.05 & 1 & 1 \\
-0.5 & 1.05 & 0 & 2 \\
0.5 & 1.05 & 0 & 3 \\
\end{tabularx}
\end{table}

\newpage
\onecolumn
\section{}\label{Appendix:G}
Full test schedule of run tests, Table~\ref{tab:TestSchedule}.

\begingroup
\scriptsize
\setlength{\tabcolsep}{1pt} 

\begin{longtable}{@{}%
p{3.8cm}
>{\centering\arraybackslash}p{0.9cm}
>{\centering\arraybackslash}p{0.9cm}
>{\centering\arraybackslash}p{1.1cm}
p{1.45cm}
p{1.25cm}
p{1.1cm}
p{1.2cm}
@{}}
\caption{Overview of benchmark configurations used in the MDBD study.}
\label{tab:TestSchedule}\\
\hline
\makecell[l]{\textbf{Name}} &
\makecell{\textbf{No.}\\\textbf{Obj.}} &
\makecell{\textbf{No.}\\\textbf{Sph.}} &
\makecell{\textbf{No.}\\\textbf{Ctrl.}} &
\makecell[l]{\textbf{Routing}\\\textbf{obj. func.}} &
\makecell[l]{\textbf{Constraint}\\\textbf{form}} &
\textbf{Type} &
\textbf{Algorithm} \\
\hline
\endfirsthead

\multicolumn{8}{c}{\tablename\ \thetable\ -- \textit{continued from previous page}} \\
\hline
\makecell[l]{\textbf{Name}} &
\makecell{\textbf{No.}\\\textbf{Obj.}} &
\makecell{\textbf{No.}\\\textbf{Sph.}} &
\makecell{\textbf{No.}\\\textbf{Ctrl.}} &
\makecell[l]{\textbf{Routing}\\\textbf{obj. func.}} &
\makecell[l]{\textbf{Constraint}\\\textbf{form}} &
\textbf{Type} &
\textbf{Algorithm} \\
\hline
\endhead

\hline \multicolumn{8}{r}{\textit{Continued on next page}} \\
\endfoot

\hline
\endlastfoot

B BH x0 O2 20 C2 Q A C N & 2 & 20 & 2 & Quadratic & Absolute & Cuboid & Nested \\
B BH x0 O2 30 C2 Q A C N & 2 & 30 & 2 & Quadratic & Absolute & Cuboid & Nested \\
B BH x0 O2 40 C2 Q A C N & 2 & 40 & 2 & Quadratic & Absolute & Cuboid & Nested \\
B BH x0 O4 20 C2 Q A C N & 4 & 20 & 2 & Quadratic & Absolute & Cuboid & Nested \\
B BH x0 O4 30 C2 Q A C N & 4 & 30 & 2 & Quadratic & Absolute & Cuboid & Nested \\
B BH x0 O4 40 C2 Q A C N & 4 & 40 & 2 & Quadratic & Absolute & Cuboid & Nested \\
B BH x0 O6 20 C2 Q A C N & 6 & 20 & 2 & Quadratic & Absolute & Cuboid & Nested \\
B BH x0 O6 30 C2 Q A C N & 6 & 30 & 2 & Quadratic & Absolute & Cuboid & Nested \\
B BH x0 O6 40 C2 Q A C N & 6 & 40 & 2 & Quadratic & Absolute & Cuboid & Nested \\
B BH xopt O4 20 C2 Q A C N & 4 & 20 & 2 & Quadratic & Absolute & Cuboid & Nested \\
B BH xopt O4 20 C0 Q A C N & 4 & 20 & 0 & Quadratic & Absolute & Cuboid & Nested \\
B BH xopt O4 30 C2 Q A C N & 4 & 30 & 2 & Quadratic & Absolute & Cuboid & Nested \\
B BH xopt O4 30 C0 Q A C N & 4 & 30 & 0 & Quadratic & Absolute & Cuboid & Nested \\
B BH xopt O4 40 C2 Q A C N & 4 & 40 & 2 & Quadratic & Absolute & Cuboid & Nested \\
B BH xopt O4 40 C0 Q A C N & 4 & 40 & 0 & Quadratic & Absolute & Cuboid & Nested \\
B ES O2 20 C0 Q A C N & 2 & 20 & 0 & Quadratic & Absolute & Cuboid & Nested \\
B ES O2 20 C2 Q A C N & 2 & 20 & 2 & Quadratic & Absolute & Cuboid & Nested \\
B ES O2 20 C4 Q A C N & 2 & 20 & 4 & Quadratic & Absolute & Cuboid & Nested \\
B ES O2 30 C0 Q A C N & 2 & 30 & 0 & Quadratic & Absolute & Cuboid & Nested \\
B ES O2 30 C2 Q A C N & 2 & 30 & 2 & Quadratic & Absolute & Cuboid & Nested \\
B ES O2 30 C4 Q A C N & 2 & 30 & 4 & Quadratic & Absolute & Cuboid & Nested \\
B ES O2 40 C0 Q A C N & 2 & 40 & 0 & Quadratic & Absolute & Cuboid & Nested \\
B ES O2 40 C2 Q A C N & 2 & 40 & 2 & Quadratic & Absolute & Cuboid & Nested \\
B ES O2 40 C4 Q A C N & 2 & 40 & 4 & Quadratic & Absolute & Cuboid & Nested \\
B ES O4 20 C0 Q A C N & 4 & 20 & 0 & Quadratic & Absolute & Cuboid & Nested \\
B ES O4 20 C2 Q A C N & 4 & 20 & 2 & Quadratic & Absolute & Cuboid & Nested \\
B ES O4 20 C4 Q A C N & 4 & 20 & 4 & Quadratic & Absolute & Cuboid & Nested \\
B ES O4 30 C0 Q A C N & 4 & 30 & 0 & Quadratic & Absolute & Cuboid & Nested \\
B ES O4 30 C2 Q A C N & 4 & 30 & 2 & Quadratic & Absolute & Cuboid & Nested \\
B ES O4 30 C4 Q A C N & 4 & 30 & 4 & Quadratic & Absolute & Cuboid & Nested \\
B ES O4 40 C0 Q A C N & 4 & 40 & 0 & Quadratic & Absolute & Cuboid & Nested \\
B ES O4 40 C2 Q A C N & 4 & 40 & 2 & Quadratic & Absolute & Cuboid & Nested \\
B ES O4 40 C4 Q A C N & 4 & 40 & 4 & Quadratic & Absolute & Cuboid & Nested \\
B ES O6 20 C0 Q A C N & 6 & 20 & 0 & Quadratic & Absolute & Cuboid & Nested \\
B ES O6 20 C2 Q A C N & 6 & 20 & 2 & Quadratic & Absolute & Cuboid & Nested \\
B ES O6 20 C4 Q A C N & 6 & 20 & 4 & Quadratic & Absolute & Cuboid & Nested \\
B ES O6 30 C0 Q A C N & 6 & 30 & 0 & Quadratic & Absolute & Cuboid & Nested \\
B ES O6 30 C2 Q A C N & 6 & 30 & 2 & Quadratic & Absolute & Cuboid & Nested \\
B ES O6 30 C4 Q A C N & 6 & 30 & 4 & Quadratic & Absolute & Cuboid & Nested \\
B ES O6 40 C0 Q A C N & 6 & 40 & 0 & Quadratic & Absolute & Cuboid & Nested \\
B ES O6 40 C2 Q A C N & 6 & 40 & 2 & Quadratic & Absolute & Cuboid & Nested \\
B ES O6 40 C4 Q A C N & 6 & 40 & 4 & Quadratic & Absolute & Cuboid & Nested \\
B GA O2 20 C0 Q A C N & 2 & 20 & 0 & Quadratic & Absolute & Cuboid & Nested \\
B GA O2 20 C2 E A C N & 2 & 20 & 2 & Exponential & Absolute & Cuboid & Nested \\
B GA O2 20 C2 Q A C N & 2 & 20 & 2 & Quadratic & Absolute & Cuboid & Nested \\
B GA O2 20 C2 Q S C N & 2 & 20 & 2 & Quadratic & Softsum & Cuboid & Nested \\
B GA O2 20 C4 Q A C N & 2 & 20 & 4 & Quadratic & Absolute & Cuboid & Nested \\
B GA O2 30 C0 Q A C N & 2 & 30 & 0 & Quadratic & Absolute & Cuboid & Nested \\
B GA O2 30 C2 Q A C N & 2 & 30 & 2 & Quadratic & Absolute & Cuboid & Nested \\
B GA O2 30 C4 Q A C N & 2 & 30 & 4 & Quadratic & Absolute & Cuboid & Nested \\
B GA O2 40 C0 Q A C N & 2 & 40 & 0 & Quadratic & Absolute & Cuboid & Nested \\
B GA O2 40 C2 Q A C N & 2 & 40 & 2 & Quadratic & Absolute & Cuboid & Nested \\
B GA O2 40 C4 Q A C N & 2 & 40 & 4 & Quadratic & Absolute & Cuboid & Nested \\
B GA O4 20 C0 Q A C N & 4 & 20 & 0 & Quadratic & Absolute & Cuboid & Nested \\
B GA O4 20 C2 E A C N & 4 & 20 & 2 & Exponential & Absolute & Cuboid & Nested \\
B GA O4 20 C2 Q A C N & 4 & 20 & 2 & Quadratic & Absolute & Cuboid & Nested \\
B GA O4 20 C2 Q A L N & 4 & 20 & 2 & Quadratic & Absolute & L-shape & Nested \\
B GA O4 20 C2 Q A U N & 4 & 20 & 2 & Quadratic & Absolute & Unique & Nested \\
B GA O4 20 C2 Q S C N & 4 & 20 & 2 & Quadratic & Softsum & Cuboid & Nested \\
B GA O4 20 C4 Q A C N & 4 & 20 & 4 & Quadratic & Absolute & Cuboid & Nested \\
B GA O4 30 C0 Q A C N & 4 & 30 & 0 & Quadratic & Absolute & Cuboid & Nested \\
B GA O4 30 C2 Q A C N & 4 & 30 & 2 & Quadratic & Absolute & Cuboid & Nested \\
B GA O4 30 C4 Q A C N & 4 & 30 & 4 & Quadratic & Absolute & Cuboid & Nested \\
B GA O4 40 C0 Q A C N & 4 & 40 & 0 & Quadratic & Absolute & Cuboid & Nested \\
B GA O4 40 C2 Q A C N & 4 & 40 & 2 & Quadratic & Absolute & Cuboid & Nested \\
B GA O4 40 C4 Q A C N & 4 & 40 & 4 & Quadratic & Absolute & Cuboid & Nested \\
B GA O6 20 C0 Q A C N & 6 & 20 & 0 & Quadratic & Absolute & Cuboid & Nested \\
B GA O6 20 C2 E A C N & 6 & 20 & 2 & Exponential & Absolute & Cuboid & Nested \\
B GA O6 20 C2 Q A C N & 6 & 20 & 2 & Quadratic & Absolute & Cuboid & Nested \\
B GA O6 20 C2 Q S C N & 6 & 20 & 2 & Quadratic & Softsum & Cuboid & Nested \\
B GA O6 20 C4 Q A C N & 6 & 20 & 4 & Quadratic & Absolute & Cuboid & Nested \\
B GA O6 30 C0 Q A C N & 6 & 30 & 0 & Quadratic & Absolute & Cuboid & Nested \\
B GA O6 30 C2 Q A C N & 6 & 30 & 2 & Quadratic & Absolute & Cuboid & Nested \\
B GA O6 30 C4 Q A C N & 6 & 30 & 4 & Quadratic & Absolute & Cuboid & Nested \\
B GA O6 40 C0 Q A C N & 6 & 40 & 0 & Quadratic & Absolute & Cuboid & Nested \\
B GA O6 40 C2 Q A C N & 6 & 40 & 2 & Quadratic & Absolute & Cuboid & Nested \\
B GA O6 40 C4 Q A C N & 6 & 40 & 4 & Quadratic & Absolute & Cuboid & Nested \\
B Ra O2 20 C0 Q A C A & 2 & 20 & 0 & Quadratic & Absolute & Cuboid & ATC \\
B Ra O2 20 C0 Q A C N & 2 & 20 & 0 & Quadratic & Absolute & Cuboid & Nested \\
B Ra O2 20 C0 Q A C S & 2 & 20 & 0 & Quadratic & Absolute & Cuboid & SOI \\
B Ra O2 20 C0 Q A L N & 2 & 20 & 0 & Quadratic & Absolute & L-Shape & Nested \\
B Ra O2 20 C2 Q A C N & 2 & 20 & 2 & Quadratic & Absolute & Cuboid & Nested \\
B Ra O2 20 C4 Q A C N & 2 & 20 & 4 & Quadratic & Absolute & Cuboid & Nested \\
B Ra O2 30 C0 Q A C N & 2 & 30 & 0 & Quadratic & Absolute & Cuboid & Nested \\
B Ra O2 30 C0 Q A L N & 2 & 30 & 0 & Quadratic & Absolute & L-Shape & Nested \\
B Ra O2 30 C2 Q A C N & 2 & 30 & 2 & Quadratic & Absolute & Cuboid & Nested \\
B Ra O2 30 C4 Q A C N & 2 & 30 & 4 & Quadratic & Absolute & Cuboid & Nested \\
B Ra O2 40 C0 Q A C N & 2 & 40 & 0 & Quadratic & Absolute & Cuboid & Nested \\
B Ra O2 40 C0 Q A L N & 2 & 40 & 0 & Quadratic & Absolute & L-Shape & Nested \\
B Ra O2 40 C2 Q A C N & 2 & 40 & 2 & Quadratic & Absolute & Cuboid & Nested \\
B Ra O2 40 C4 Q A C N & 2 & 40 & 4 & Quadratic & Absolute & Cuboid & Nested \\
B Ra O4 20 C0 Q A C A & 4 & 20 & 0 & Quadratic & Absolute & Cuboid & ATC \\
B Ra O4 20 C0 Q A C N & 4 & 20 & 0 & Quadratic & Absolute & Cuboid & Nested \\
B Ra O4 20 C0 Q A C S & 4 & 20 & 0 & Quadratic & Absolute & Cuboid & SOI \\
B Ra O4 20 C0 Q A L N & 4 & 20 & 0 & Quadratic & Absolute & L-Shape & Nested \\
B Ra O4 20 C2 Q A C N & 4 & 20 & 2 & Quadratic & Absolute & Cuboid & Nested \\
B Ra O4 20 C4 Q A C N & 4 & 20 & 4 & Quadratic & Absolute & Cuboid & Nested \\
B Ra O4 30 C0 Q A C N & 4 & 30 & 0 & Quadratic & Absolute & Cuboid & Nested \\
B Ra O4 30 C0 Q A L N & 4 & 30 & 0 & Quadratic & Absolute & L-Shape & Nested \\
B Ra O4 30 C2 Q A C N & 4 & 30 & 2 & Quadratic & Absolute & Cuboid & Nested \\
B Ra O4 30 C4 Q A C N & 4 & 30 & 4 & Quadratic & Absolute & Cuboid & Nested \\
B Ra O4 40 C0 Q A C N & 4 & 40 & 0 & Quadratic & Absolute & Cuboid & Nested \\
B Ra O4 40 C0 Q A L N & 4 & 40 & 0 & Quadratic & Absolute & L-Shape & Nested \\
B Ra O4 40 C2 Q A C N & 4 & 40 & 2 & Quadratic & Absolute & Cuboid & Nested \\
B Ra O4 40 C4 Q A C N & 4 & 40 & 4 & Quadratic & Absolute & Cuboid & Nested \\
B Ra O6 20 C0 Q A C N & 6 & 20 & 0 & Quadratic & Absolute & Cuboid & Nested \\
B Ra O6 20 C0 Q A L N & 6 & 20 & 0 & Quadratic & Absolute & L-Shape & Nested \\
B Ra O6 20 C2 Q A C N & 6 & 20 & 2 & Quadratic & Absolute & Cuboid & Nested \\
B Ra O6 20 C4 Q A C N & 6 & 20 & 4 & Quadratic & Absolute & Cuboid & Nested \\
B Ra O6 30 C0 Q A C N & 6 & 30 & 0 & Quadratic & Absolute & Cuboid & Nested \\
B Ra O6 30 C0 Q A L N & 6 & 30 & 0 & Quadratic & Absolute & L-Shape & Nested \\
B Ra O6 30 C2 Q A C N & 6 & 30 & 2 & Quadratic & Absolute & Cuboid & Nested \\
B Ra O6 30 C4 Q A C N & 6 & 30 & 4 & Quadratic & Absolute & Cuboid & Nested \\
B Ra O6 40 C0 Q A C N & 6 & 40 & 0 & Quadratic & Absolute & Cuboid & Nested \\
B Ra O6 40 C0 Q A L N & 6 & 40 & 0 & Quadratic & Absolute & L-Shape & Nested \\
B Ra O6 40 C2 Q A C N & 6 & 40 & 2 & Quadratic & Absolute & Cuboid & Nested \\
B Ra O6 40 C4 Q A C N & 6 & 40 & 4 & Quadratic & Absolute & Cuboid & Nested \\

\end{longtable}
\endgroup

\newpage
\section{}\label{Appendix:H}
Visual results ATC, SOI and Nested.

\begin{figure}[h!]
    \centering
    \includegraphics[width=0.5\linewidth]{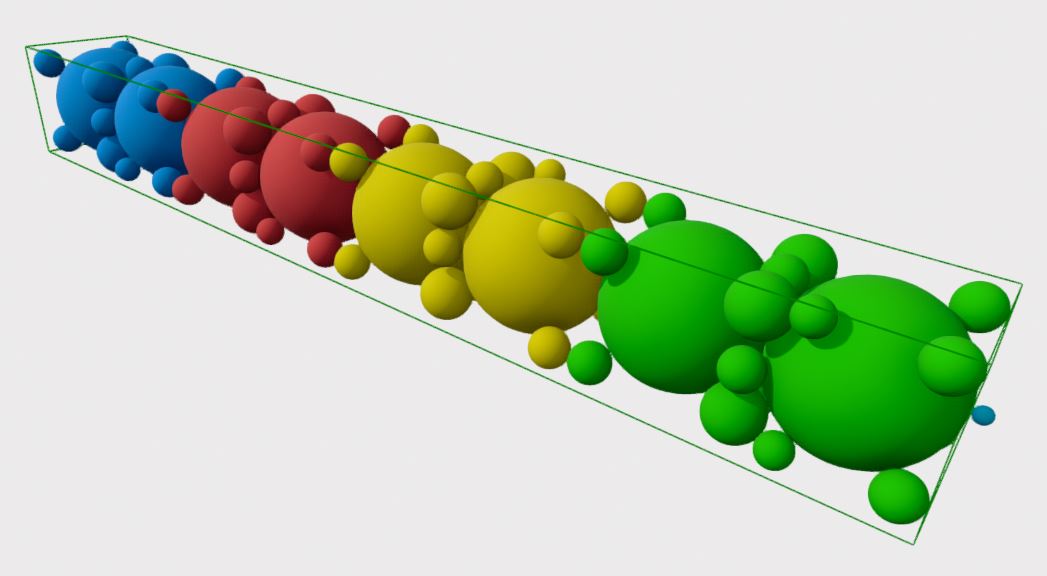}
    \caption{The visual results of the nested algorithm with 4 objects, no routing and 20 spheres}
    \label{fig:ResultNested}
\end{figure}

\begin{figure}[h!]
    \centering
    \includegraphics[width=0.5\linewidth]{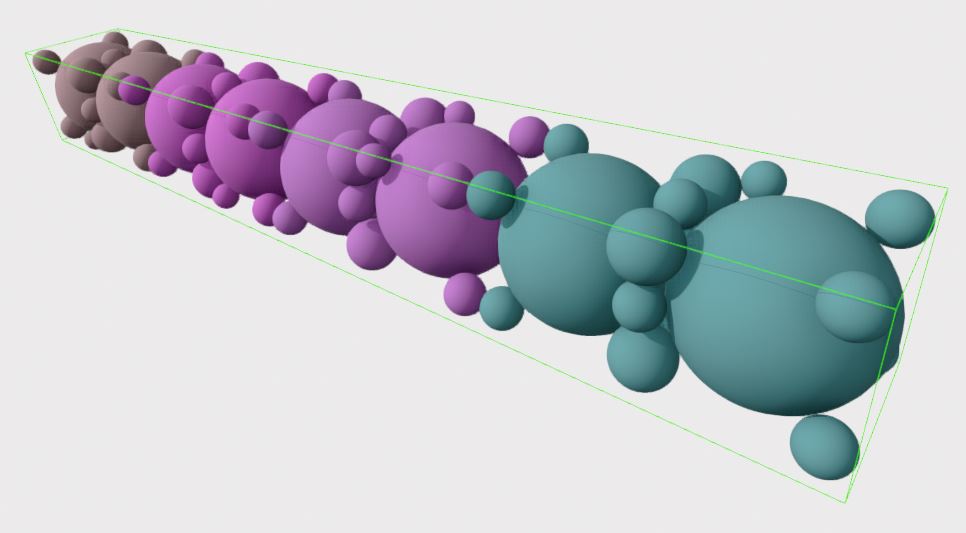}
    \caption{The visual results of the SOI algorithm with 4 objects, no routing and 20 spheres}
    \label{fig:ResultSOI}
\end{figure}

\begin{figure}[h!]
    \centering
    \includegraphics[width=0.5\linewidth]{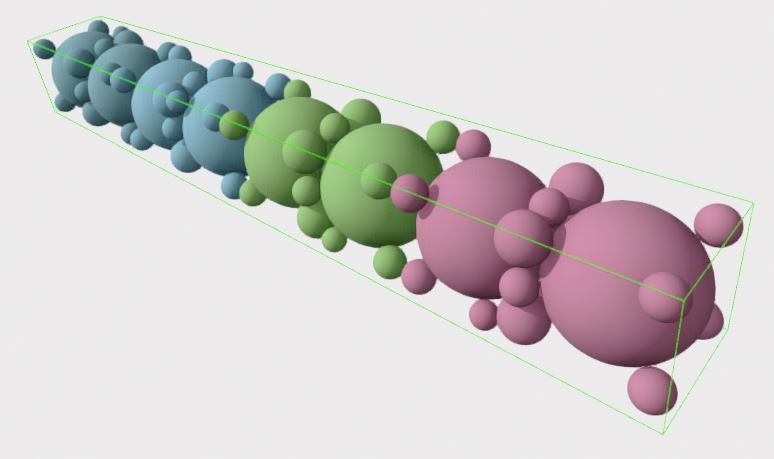}
    \caption{The visual results of the ATC algorithm with 4 objects, no routing and 20 spheres}
    \label{fig:ResultATC}
\end{figure}
\newpage
\section{}\label{Appendix:Z}
\begin{table}[h!]
\centering
\caption{Variables and notation (ordered by symbol) — Part 1}
\footnotesize
\setlength{\tabcolsep}{5pt}
\renewcommand{\arraystretch}{1.15}
\begin{tabularx}{\columnwidth}{@{}l l X@{}}
\textbf{Symbol} & \textbf{Type} & \textbf{Meaning / Range} \\
\hline
\(A_i\) & set & Object \(i\) represented as a set of spheres: \(A_i=\{b_{i,1},\dots,b_{i,n_{\mathrm{b},i}}\}\). \\
\(\mathcal{C}\) & set & Checked routing–routing segment-pair set defined in~\eqref{eq:SetS}. \\
\(\mathbb{F}_{A_i}\) & frame & Object frame attached to \(A_i\). \\
\(\mathbb{F}_\mathbb{W}\) & frame & Cartesian frame attached to \(\mathbb{W}\). \\
\(K_L\) & int & Number of segments in route \(L\) (\(K_L=n_{\mathrm{cp},L}+1\)). \\
\(L\) & index & Route index. \\
\(N_{\mathrm{pairs}}\) & int & Number of sphere–pair constraints across all objects. \\
\(N_{\mathrm{r\mbox{-}o}}\) & int & Number of routing–object interference constraints. \\
\(N_{\mathrm{r\mbox{-}r}}\) & int & Number of routing–routing interference constraints (\(|\mathcal{C}|\)). \\
\(S_{\alpha}(\cdot)\) & operator & Boltzmann (soft) operator; smooth approx. of \(\max/\min\) depending on sign of \(\alpha\). \\
\(\mathbb{W}\) & set & Workspace in \(\mathbb{R}^{n_\mathrm{d}}\). \\
\(\alpha\) & scalar & Soft-operator sharpness parameter. \\
\(b_{i,\mu}\) & sphere & Sphere \(\mu\) of object \(i\). \\
\(\mathbf{c}_{L,k}\) & vector \(\in\mathbb{R}^3\) & Control point \(k\) of route \(L\). \\
\(d^{\,\mathrm{obj\mbox{-}obj}}_{i,\mu,j,\nu}\) & scalar & Clearance between spheres \(b_{i,\mu}\) and \(b_{j,\nu}\) (Eq.~\eqref{eq:clearance_objobj}). \\
\(d^{\,\mathrm{route\mbox{-}obj}}_{L,m,i,\mu}\) & scalar & Clearance between routing segment \((L,m)\) and sphere \(b_{i,\mu}\). \\
\(d^{\,\mathrm{route\mbox{-}route}}_{L,m,L',\eta}\) & scalar & Clearance between routing segments \((L,m)\) and \((L',\eta)\). \\
\(f_\mathrm{re}\) & scalar & Exponential routing-length objective. \\
\(f_\mathrm{rq}\) & scalar & Squared routing-length objective \eqref{eq:frl}. \\
\(f_\mathrm{v}\) & scalar & AABB volume using objects only. \\
\(f_\mathrm{vr}\) & scalar & AABB volume using objects and routing points. \\
\(\mathbf{g}^{\mathrm{obj\mbox{-}obj}}\) & vector & Stacked object–object non-overlap constraints. \\
\(\mathbf{g}^{\mathrm{route\mbox{-}obj}}\) & vector & Stacked routing–object non-overlap constraints. \\
\end{tabularx}
\end{table}

\begin{table}[h!]
\centering
\caption{Variables and notation (ordered by symbol) — Part 2}
\footnotesize
\setlength{\tabcolsep}{5pt}
\renewcommand{\arraystretch}{1.15}
\begin{tabularx}{\columnwidth}{@{}l l X@{}}
\textbf{Symbol} & \textbf{Type} & \textbf{Meaning / Range} \\
\hline
\(\mathbf{g}^{\mathrm{route\mbox{-}route}}\) & vector & Stacked routing–routing non-overlap constraints. \\
\(g_\mathrm{soft}^{\mathrm{obj\mbox{-}obj}}\) & scalar & Soft aggregate of \(\mathbf{g}^{\mathrm{obj\mbox{-}obj}}\). \\
\(g_\mathrm{soft}^{\mathrm{route\mbox{-}obj}}\) & scalar & Soft aggregate of \(\mathbf{g}^{\mathrm{route\mbox{-}obj}}\). \\
\(g_\mathrm{soft}^{\mathrm{route\mbox{-}route}}\) & scalar & Soft aggregate of \(\mathbf{g}^{\mathrm{route\mbox{-}route}}\). \\
\(i,j\) & indices & Object indices (\(1\le i<j\le n_\mathrm{obj}\)). \\
\(k\) & index & Control-point index. \\
\(\ell\) & index & Port index on object \(i\) (\(1,\dots,n_{\varphi,i}\)). \\
\(m,\eta\) & indices & Routing segment indices. \\
\(n_\mathrm{d}\) & int & Workspace dimension (here \(=3\)). \\
\(n_\mathrm{obj}\) & int & Number of rigid objects. \\
\(n_{\mathrm{b},i}\) & int & Number of spheres representing object \(i\). \\
\(n_{\mathrm{cp},L}\) & int & Number of control points of route \(L\). \\
\(n_\mathrm{prob}\) & int & Problem index in \(\{1,2,3,4\}\). \\
\(n_\mathrm{var}\) & int & Total number of design variables \(=6(n_\mathrm{obj}-1)+3\sum_L n_{\mathrm{cp},L}\). \\
\(\mathbf{p}^{A_i}\) & vector \(\in\mathbb{R}^3\) & Point expressed in object frame \(\mathbb{F}_{A_i}\). \\
\(\mathbf{p}^{\mathbb{W}}\) & vector \(\in\mathbb{R}^3\) & Transformed point in workspace frame \(\mathbb{F}_\mathbb{W}\). \\
\(\mathbf{p}^{\mathbb{W}}_{\mathrm{b}_{i,\mu}}\) & vector \(\in\mathbb{R}^3\) & Center of sphere \(b_{i,\mu}\) in workspace frame. \\
\(\mathbf{p}_{\mathrm{proj},L,m}\) & vector \(\in\mathbb{R}^3\) & Closest point on routing segment \((L,m)\) to a given sphere center. \\
\(\mathbf{q}_{L,k}\) & vector \(\in\mathbb{R}^3\) & Route node: start/end port or control point. \\
\(r_{\mathrm{b}_{i,\mu}}\) & scalar & Radius of sphere \(b_{i,\mu}\). \\
\(r_r\) & scalar & Routing-tube radius. \\
\(\mathbf{R}_i\) & matrix \(\in\mathbb{R}^{3\times 3}\) & Rotation from \(\mathbb{F}_{A_i}\) to \(\mathbb{F}_\mathbb{W}\) (RPY composition). \\
\(\mathbf{t}_i\) & vector \(\in\mathbb{R}^3\) & Translation of object \(i\) in \(\mathbb{F}_\mathbb{W}\). \\
\(\varphi_{i,\ell}\) & point & Port \(\ell\) on object \(i\). \\
\(w_\mathrm{v},w_\mathrm{vr},w_\mathrm{rq},w_\mathrm{re}\) & scalars & Weights for the objective functions. \\
\(\mathbf{x}\) & vector & Stacked design variables \([\mathbf{x}_2^\top,\dots,\mathbf{x}_{n_\mathrm{obj}}^\top,\mathbf{c}_{1,1}^\top,\dots]^\top\). \\
\(\mathbf{x}_{\mathrm{o}i}\) & vector \(\in\mathbb{R}^6\) & Pose of object \(i\): RPY \((x_{\theta,i},x_{\alpha,i},x_{\beta,i})\) and translation \((x_{x,i},x_{y,i},x_{z,i})\). \\
\(\mu,\nu\) & indices & Sphere indices for objects \(i\) and \(j\). \\
\(\zeta\) & Scalar & Stabilization weight (small regularization term) applied to the subsystem objective\\
\end{tabularx}
\end{table}

\end{document}